\documentclass{article}

\usepackage{arxiv}
\usepackage{authblk}
\usepackage{subfig}
\usepackage{graphicx}
\usepackage{epstopdf, epsfig}

\usepackage[export]{adjustbox}
\usepackage{amsmath,amssymb}
\usepackage{booktabs}
\usepackage[numbers,sort&compress]{natbib}
\usepackage{longtable,tabularx}
\usepackage{xcolor}
\usepackage{cleveref}
\usepackage{makecell}
\usepackage{tablefootnote} 

\usepackage{url}

\crefformat{section}{\S#2#1#3}

\newcommand{\newpart}[1]{{\color{magenta}}}

\newcommand{\SRsub}{\mathrm{SR}}
\newcommand{\LRsub}{\mathrm{LR}}

\newcommand{\grad}{\nabla}

\title{Dynamic mixed turbulence modeling using a super-resolution generative adversarial approach} 

\author[1]{L. Nista}
\author[2]{C. D. K. Schumann}
\author[3]{T. Grenga}
\author[4]{J. F. MacArt}
\author[5]{A. Attili}
\author[1]{H. Pitsch}

\affil[1]{Institute for Combustion Technology, RWTH Aachen University, Aachen, 52056, Germany}
\affil[2]{Department of Engineering, University of Cambridge, Cambridge, CB2 1PZ, United Kingdom}
\affil[3]{Department of Aeronautics and Astronautics, Faculty of Engineering and Physical Sciences, University of Southampton, Southampton, SO17 1BJ, United Kingdom}
\affil[4]{Department of Aerospace and Mechanical Engineering, University of Notre Dame, Notre Dame, IN 46556, USA}
\affil[5]{School of Engineering, Institute for Multiscale Thermofluids, University of Edinburgh, Edinburgh, EH9 3FD, United Kingdom}

\begin{document}

\maketitle

\begin{abstract} 
A dynamic mixed super-resolution model (DMSRM) for large-eddy simulations (LESs) is proposed, which combines the traditional dynamic mixed model (DMM) formulation with the generation of super-resolved velocity fields from which the subfilter-scale (SFS) stress tensor can be computed. A data-driven super-resolution generative adversarial network (SR-GAN) is employed to upsample the grid-filtered velocity fields by a factor of two, enabling the evaluation of both scale-similarity and the dynamic Smagorinsky contributions.

\textit{A priori} analyses of forced homogeneous isotropic turbulence (HIT) show that the SR-GAN accurately reconstructs flow features and generalizes well across different filter sizes and higher Reynolds numbers within the same flow configuration, even for unseen input fields. The DMSRM reproduces SFS stresses and energy dissipation more accurately than the traditional DMM. \textit{A posteriori} LES calculations further confirm that DMSRM predicts the energy spectrum and intermittency more accurately than DMM, even for different LES grid-scale resolutions and higher Reynolds numbers than those used for training within the same forced HIT flow configuration. Unlike DMM, DMSRM yields realistic backscatter and physically consistent SFS energy dissipation. 

These improvements arise from the physically accurate super-resolved fields generated by the SR-GAN, from which SFS stresses are directly computed. The result is a closure that accurately reproduces stress magnitudes and dissipation while reducing reliance on additional dissipation from the dynamic term. The DMSRM formulation achieves a balance of physical fidelity, robustness, and computational efficiency, offering a promising alternative to traditional DMMs for turbulence LES modeling. 
\end{abstract}

\section{Introduction}
\label{sec:introduction}

Direct numerical simulation (DNS) resolves all time and length scales of turbulent flows but faces prohibitive computational costs for engineering applications. To address this, large-eddy simulation (LES) solves spatially filtered equations, resulting in significant cost reduction \citep{piomelli1996large, durbin2018some}. The LES flow field is decomposed into resolved and subfilter-scale (SFS) components, and the filtered governing equations are solved with the SFS terms requiring closure. Accurate closure models are therefore essential for capturing the subfilter-scale influence on the resolved scales \citep{sagaut2006large}. While we focus on models for the SFS stress tensor, analogous challenges exist for other SFS terms, especially when reacting flows are considered \citep{pitschReviewLES}.

Various approaches have been proposed to model the SFS contributions in incompressible flows, which are broadly classified into functional and structural models. Functional models, such as the widely used dynamic Smagorinsky model (DSM) \citep{germano1991dynamic}, capture the mean dissipative effect of SFS motions on the resolved scales, often through an eddy-viscosity formulation. However, correctly predicting the mean dissipation might be insufficient to reproduce the correct flow statistics, and \textit{a priori} analyses demonstrated that the DSM's modeled and exact SFS stress tensors do not correlate well \citep{meneveau2000scale}. Conversely, structural models, such as the original scale-similarity model (SSM) \citep{bardina1983improved}, aim to improve the fidelity of SFS stress tensors by leveraging the assumption of scale similarity. The velocity field at scales below LES grid resolution is postulated to be similar to that at scales above, and an additional test filter is applied at the resolved scale to capture the structural characteristics of SFS turbulence \citep{liu1994properties}. \textit{A priori} studies have shown that the stresses modeled with the SSM are more similar to the exact SFS stress than those with the DSM; however, they generally do not provide sufficient dissipation \textit{a posteriori}, which can lead to numerical instability.

A straightforward approach for developing closure models that simultaneously possess (a) favorable structural properties, and (b) enough mean energy dissipation, is to combine functional and structural models into the so-called dynamic mixed models (DMMs) \citep{zang_mixed}. In DMMs, the structural scale-similarity component provides an accurate representation of the SFS stress tensor, while the functional dissipative component ensures stability \citep{sagaut2006large}. 

Since the physical accuracy of DMMs strongly depends on the fidelity of the scale-similarity contribution, several improvements to the original SSM formulation have been proposed, specifically regarding the choices for the test filter kernel and width, and the number of dynamic model coefficients \citep{anderson1999effects, vreman1994formulation, salvetti1995priori}. Additionally, a number of approaches have been proposed as alternatives to traditional similarity-based models, which postulate a representation of the SFS velocity field, from which the SFS stresses can be reconstructed. These methods differ in complexity and cost, often introducing additional assumptions or requiring auxiliary equations. For example, Scotti \textit{et al.}~\cite{fractalInterpolation} proposed a method based on fractal representations of turbulence. This model is computationally inexpensive but neglects the local flow dynamics and requires an additional transport equation. McDonough \textit{et al.}~\cite{mcdonough1998data} introduced a chaotic map model that mimics turbulent fluctuations via a low-dimensional chaotic dynamical system. Though simple and efficient, this method requires the specification of suitable dynamics, which is non-trivial for complex flows. An approach proposed by Domaradzki \textit{et al.}~\cite{domaradzki2002large} involves explicit computation of SFS fluctuations by solving a simplified advection equation. This technique is cost-effective but requires an approximate inverse filter, which is challenging to specify \citep{sagaut2006large}. More recently, data-driven deep learning (DL) approaches have also been explored, particularly through super-resolution (SR) frameworks. Low-resolution (LR) LES fields are upsampled to high-resolution (HR) representations on finer meshes using data-driven SR frameworks, which recover unresolved flow features by learning multi-scale structures and intermittency directly from high-fidelity data  \citep{fukami2023super, kim2021unsupervised}. In particular, generative convolutional models such as generative adversarial networks (GANs) have demonstrated DNS-level reconstruction, strong \textit{a priori} generalization to unseen flow conditions, and significantly better accuracy than traditional algebraic models in \textit{a priori} tests \citep{vinuesa2021potential, kim2021unsupervised, IHME2022101010, nista2024PRF, wu2025generalizable}.

While \textit{a priori} tests have shown remarkable model-fitting capabilities across different flow conditions, \textit{a posteriori} LES with data-driven SR closure models (SR-LES) remain scarce, especially for large upsampling factors, i.e., for large LES-to-DNS filter ratios, and high Reynolds number flows. This limitation arises because the cost of training and evaluating SR models scales with the upsampling factor and is driven by the increased training dataset size, complexity of learning, and number of trainable parameters. Even more importantly, the memory and inference time requirements of DNS-level SR pose significant barriers to scalability \citep{10.1007/978-3-030-44584-3_43}. Furthermore, SR-LES imposes restrictions on domain decomposition due to convolutional boundary effects, which limit parallel scalability and increase computational cost \citep{Nista2024_parallel}. As a result, employing SR-LES for full-scale (DNS-like) SFS reconstruction is impractical even for moderate Reynolds number flows.

Nevertheless, SR-based methods may remain promising for SFS modeling when combined with existing LES model formulations. A known limitation of scale similarity and mixed models is that the SFS stress tensor is derived from the grid-filtered velocity field, which is typically numerically inaccurate. In addition, a test filter is usually applied, which can further increase errors. Even if the grid-filtered velocity field is considered physically accurate, the arbitrary choice of test filter kernel and size may reduce model fidelity. In contrast, SR-GANs leverage high-fidelity DNS data and physics-informed adversarial training to generate super-resolved velocity fields that are consistent with actual turbulence dynamics \citep{nista2024PRF}; these fields are then used to evaluate the SFS terms, potentially improving predictive accuracy even when applied to imperfect grid-filtered velocity fields.

Based on the potential advantages outlined above, we introduce a dynamic mixed SR-based model (DMSRM) (Section~\ref{sec:dynamic_mixed_SR_SS_model}) that combines (a) the robustness and computational efficiency of DMMs, (b) SR-generated velocity fields for scale-similarity-based SFS stress computation, and (c) the physically consistent SR reconstruction capabilities of SR-GANs. By leveraging SR-GANs (Section~\ref{sec:architecture}), the proposed formulation avoids arbitrary test filters by explicitly reconstructing super-resolved velocity fields at a slightly increased resolution compared to the LES grid. This enables the use of a scale-similarity-like formulation with a modest upsampling ratio of two, thereby avoiding the computational cost of conventional SR-based approaches. We evaluate the DMSRM using two forced homogeneous isotropic turbulence (HIT) datasets (Section~\ref{sec:dataset}), examining both its \textit{a priori} (Section~\ref{sec:apriori_sec}) and \textit{a posteriori} (Section~\ref{sec:aposteriori_sec}) performance with various training filter kernels, filter sizes, and flow conditions. Additionally, we assess the model’s numerical stability and the individual contributions of its components (Section~\ref{sec:aposteriori_stability}), as well as its computational efficiency relative to the standard DMM (Section~\ref{sec:computationalCost}). 

The overarching objective is to develop a hybrid modeling framework that leverages the capabilities of DL-based SR frameworks while maintaining feasible computational cost and preserving the stability and efficiency of DMMs. In this regard, the novelty of this work is to introduce a model formulation that operates exclusively between two LES-scale resolutions, both far from the DNS scale, and to demonstrate that a modest upsampling factor is sufficient to substantially improve the scale-similarity estimate of the DMM without incurring the prohibitive computational cost of DNS-resolution SR.

\section{LES formulation and dynamic mixed models}
\label{sec:dynamic_mixed_SR_SS_model}

The filtered momentum equation for incompressible flow is
\begin{equation}
   \frac{\partial \overline{u}_j}{\partial t} + \frac{\partial (\overline{u}_i \overline{u}_j)}{\partial x_i} = -\frac{1}{\rho} \frac{\partial \overline{p}}{\partial x_j} + \nu \frac{\partial^2 \overline{u}_j}{\partial x_i \partial x_i} - \frac{\partial \tau_{ij}}{\partial x_i} + \overline{f}_j \, \mathrm{with} \, j=1,2,3 \,,
    \label{eqn:NSequation}
\end{equation}
where $\overline{u}_j$ are the LES-filtered velocity components, the filtered pressure field $\overline{p}$ includes the isotropic part of the SFS stress tensor, $\rho$ is the density, $\nu$ is the kinematic viscosity, and $\overline{f}_{j}$ is a volumetric forcing term (see Section~\ref{sec:dataset}). We consider grid-filtered LES solutions of Eq.~\eqref{eqn:NSequation} represented on a discrete LES domain $\overline{\Omega}$ with $\overline{N}_x\times \overline{N}_y\times \overline{N}_z$ points, where $\overline{N}_i$ is the number of LES grid points in the $i^\text{th}$ direction, and $\overline{\Delta}$ is the uniform LES grid spacing. In Eq.~\eqref{eqn:NSequation}, the deviatoric SFS stress tensor is
\begin{equation}
    \tau_{ij} = \tau^{R}_{ij} - \frac{2}{3} k_r \delta_{ij}, \quad \mathrm{with} \quad  \tau^{R}_{ij} = \overline{u_i u}_j - \overline{u}_i \, \overline{u}_j \quad \mathrm{and} \quad k_r \equiv \frac{1}{2} \tau^{R}_{ii} \, ,
    \label{eqn:SFStrue}
\end{equation}
where $\tau^{R}_{ij}$ is the SFS stress tensor and $k_r$ is the residual kinetic energy. 

\subsection{Dynamic Smagorinsky model}

The (dynamic) Smagorinsky model for $\tau_{ij}$ is
\begin{equation}
    \tau^{\mathrm{DSM}}_{ij} = -2 (C_s \overline{\Delta})^2 |\overline{S}|\overline{S}_{ij} \, ,
    \label{eqn:DSM}
\end{equation}
where in the dynamic variant, the coefficient $C^2_s$ is computed using the procedure of Germano \textit{et al.}~\cite{germano1991dynamic}. To ensure numerical stability, spatial averaging is applied to the dynamically computed coefficient as proposed by Lilly~\citep{lilly1992proposed}. In Eq.~\eqref{eqn:DSM}, the filtered strain-rate tensor $\overline{S}_{ij}$ and its magnitude $|\overline{S}|$ are 
\begin{equation}
    \overline{S}_{ij} = \frac{1}{2} \left(\frac{\partial \overline{u}_{i}}{\partial x_j} + \frac{\partial \overline{u}_{j}}{\partial x_i} \right) , \qquad
    |\overline{S}| = (2\,\overline{S}_{ij} \overline{S}_{ij})^{1/2}  \, .
    \label{eqn:strain_rate}
\end{equation}
The dynamic procedure operates by applying a low-pass \emph{test filter} of width $2\overline{\Delta}$ to the velocity field and invoking the scale-similarity hypothesis over the inter-scale range (i.e., for scales between $\overline{\Delta}$ and $2\overline{\Delta}$). The dynamic Smagorinsky model is a component of both the dynamic mixed model and our proposed dynamic mixed super-resolution-based model, though their inter-scale ranges are different. We discuss these two models in the subsequent subsections.

\subsection{Dynamic mixed model}
The dynamic mixed model combines the dynamic Smagorinsky model and the scale-similarity model as 
\begin{align}
    \tau^{\mathrm{DMM}}_{ij} &= \tau^{\mathrm{DSM}}_{ij} + \tau^{\mathrm{SSM}}_{ij} = \underbrace{-2 (C_{d} \overline{\Delta})^2 |\overline{S}| \overline{S}_{ij}}_{\text{DSM $\rightarrow$ DMM}} + \underbrace{C_{L} \tau^{\mathrm{SSM}}_{ij}}_{\text{SSM}} \label{eqn:DMM} \, ,
\end{align}
where the scale-similarity coefficient is $C_L=1$, and a dynamic procedure is applied to obtain $C^2_d$ (\cite{anderson1999effects}; see Appendix~\ref{sec:DMM_Cd_definition}). Here, the DSM contribution to the DMM is denoted ``DSM $\rightarrow$ DMM.'' The scale-similarity term is
\begin{equation}
    \tau^{\mathrm{SSM}}_{ij} = \widehat{\overline{u}_i \overline{u}}_j - \widehat{\overline{u}}_i \widehat{\overline{u}}_j,
    \label{eqn:SSM}
\end{equation}
where $\widehat{\cdot}$ is the test-filter operator---here, a spectral filter with kernel width $2\overline{\Delta}$ (see table~\ref{tab:filtering_description} and Appendix~\ref{sec:appendinx_filters}). 

In constructing Eq.~\eqref{eqn:DMM}, the single-parameter dynamic formulation is preferred over the two-parameter approach, as the two-parameter dynamic procedure tends to heavily favor the scale-similarity contribution due to its higher correlation with the DNS-evaluated SFS stress. This often results in a model that is insufficiently dissipative. The dynamic coefficients can be decoupled \citep{Morinishi_coupledDMM}, but this increases computational cost and is not considered here. The definition of the SSM in Eq.~\eqref{eqn:SSM} is based on the improved version of Sarghini \textit{et al.}~\cite{sarghini1999scale}, rather than the original formulation of Bardina~\cite{bardina1983improved}. An overview of the various DMM formulations is provided in \cite{meneveau2000scale, sagaut2006large} for different filter kernels, filter sizes, and dynamic coefficient evaluation methods.

\subsection{Dynamic mixed SR-based model}

The dynamic mixed super-resolution-based model is similar in spirit to the dynamic mixed model but has one principal difference: rather than evaluating the scale-similarity term and the dynamic Smagorinsky contribution for scales coarser than the LES grid, it evaluates them on finer scales, i.e., for the energy content between the LES grid-scale solution and its super-resolved counterpart. 

Let $u_i^*$ be a super-resolved velocity field, obtained by upsampling $\overline{u}_i$ using an SR-GAN model (see Section~\ref{sec:architecture}), on an auxiliary mesh $\Omega^*$ with $2\overline{N}_x\times 2\overline{N}_y\times 2\overline{N}_z$ points, and let $\Delta^* = \overline{\Delta}/2$ be the uniform mesh spacing on $\Omega^*$. While we use an upsampling ratio of two to minimize computational cost and memory requirements, in principle, any upsampling ratio of the form $2^n$, where $n \in \mathbb{N}$ is the number of upsampling layers (see Section~\ref{sec:architecture}), could be used. In the DMSRM context, we use $\widehat{\cdot}$ to denote a spectral filter kernel of width $2\Delta^*=\overline{\Delta}$ (i.e., filtering SR-scale fields to the LES-grid scale) and $\widetilde{\cdot}$ to denote a spectral filter kernel of width $4\Delta^*=2\overline{\Delta}$ (i.e., filtering SR-scale fields to twice the LES-grid scale). We also use  $\check{\cdot}$ to denote a discrete downsampling operator to restrict $\Omega^*$ fields to $\overline{\Omega}$. Table~\ref{tab:filtering_description} summarizes these operators and the domains on which they operate.
\begin{table}[!ht]
    \centering
    \begin{tabular}{c c c}
    \toprule
     Quantity & Description & Defined on \\
     \midrule
     $\overline{u}_{i}$ & Velocity field at (LES) grid-filter scale & $\overline{\Omega}$ \\
     $\overline{\mathcal{Q}}$ & Quantity evaluated using $\overline{u}_{i}$ & $\overline{\Omega}$ \\
     $u^*_{i}$ & SR-GAN reconstructed velocity field from $\overline{u}_{i}$ & $\Omega^*$ \\
     $\mathcal{Q}^*$ & Quantity evaluated using ${u}^*_{i}$ & $\Omega^*$ \\
     \\
     \midrule
     Operator (model) & Filter kernel & Mapping \\
     \midrule
      $\widehat{\cdot}$ \, (DMM) & Spectral,  $2\overline{\Delta}$  & $\overline{\Omega}$ $\rightarrow$ $\overline{\Omega}$ \\
      $\widetilde{\cdot}$ \, (DMM) & Spectral, $4\overline{\Delta}$ & $\overline{\Omega}$ $\rightarrow$ $\overline{\Omega}$   \\
      \midrule
      $\widehat{\cdot}$ \, (DMSRM) & Spectral, $2\Delta^*=\overline{\Delta}$ & $\Omega^*$ $\rightarrow$ $\Omega^*$ \\
      $\widetilde{\cdot}$ \, (DMSRM) & Spectral, $4\Delta^*=2\overline{\Delta}$ & $\Omega^*$ $\rightarrow$ $\Omega^*$   \\
      $\check{\cdot}$ \, (DMSRM) & -- (Downsampling) & $\Omega^*$ $\rightarrow$ $\overline{\Omega}$ (factor: 2) \\
      \bottomrule
    \end{tabular}
    \caption{Overview of quantities, notations, and operators employed by the DMSRM or DMM. The ``defined on'' column indicates the discrete domain on which each quantity is defined, where $\overline{\Omega}$ and $\Omega^*$ indicate the discrete LES and super-resolved domains, respectively. For operators, the ``mapping'' column specifies the input and output domains for each operator. The definitions of the spectral filter kernel and the downsampling operator are reported in Appendix~\ref{sec:appendinx_filters}.
    }
    \label{tab:filtering_description}
\end{table}

The DMSRM then employs a SR-scale term
\begin{equation}
     \tau^{*}_{ij} = -2 C_{d}^2 (\Delta^{*})^{2} |S^*| S^*_{ij} + (\widehat{u^*_i \, u^*_j} - \widehat{u^*_i} \, \widehat{u^*_j}) \, ,
    \label{eqn:SRSSmodel_SRlevel}
\end{equation}
where  $S^*_{ij}$ is the SR-scale strain-rate tensor with magnitude $|S^*| = (2\,S^*_{ij} S^*_{ij})^{1/2}$, as well as a ``test-filtered SR-scale'' (i.e., LES-scale) term
\begin{equation}
    T_{ij} = -2 C_{d}^2 (2\Delta^*)^2 |\widehat{{S^*}}|  \widehat{S^*_{ij}} + (\widetilde{\widehat{u^*_i} \,  \widehat{u^*_j}} - \widetilde{\widehat{u^*_i}} \, \widetilde{\widehat{u^*_j}}),
    \label{eqn:SRSSmodel_gridlevel}
\end{equation}
which is needed to evaluate $C^2_d$ dynamically. It is important to note that both Eqs.~\eqref{eqn:SRSSmodel_SRlevel} and \eqref{eqn:SRSSmodel_gridlevel} are evaluated on $\Omega^*$. At the SR scale, the Germano identity \citep{germano1991dynamic} is
\begin{equation}
    \mathcal{L}_{ij} = (\widehat{{u}^*_i \, u^*_j} - \widehat{u^*_i} \, \widehat{u^*_j}) = T_{ij} - \widehat{\tau}^*_{ij} \, .
    \label{eqn:Germano_identity}
\end{equation}
Substituting Eq.~\eqref{eqn:SRSSmodel_SRlevel} and Eq.~\eqref{eqn:SRSSmodel_gridlevel} into Eq.~\eqref{eqn:Germano_identity} yields
\begin{align}
    \mathcal{L}_{ij} &= C_{d}^2 \mathcal{M}_{ij} + \mathcal{N}_{ij},  \nonumber  \\ 
    \mathcal{M}_{ij} &= -2(\Delta^*)^2 (4 |\widehat{S^*}| \widehat{{S}^*_{ij}} - \widehat{|S^*| \, S^*_{ij}}), \nonumber \\
   \mathcal{N}_{ij} &= (\widetilde{\widehat{u^*_i} \,  \widehat{u^*_j}} - \widetilde{\widehat{u^*_i}} \, \widetilde{\widehat{u^*_j}}) - (\widehat{\widehat{{u}^*_i \, u^*_j}} - \widehat{\widehat{u^*_i} \, \widehat{u^*_j}}) \, .
    \label{eqn:M_IJ_L_IJ_N_IJ}
\end{align}
The model coefficient $C_{d}^2$ is evaluated dynamically as
\begin{equation}
C_d^2 = \frac{\langle \mathcal{L}_{ij} \mathcal{M}_{ij}\rangle - \langle \mathcal{N}_{ij} \mathcal{M}_{ij}\rangle}{\langle \mathcal{M}_{ij} \mathcal{M}_{ij}\rangle},
\label{eqn:C_d}
\end{equation}
where $\langle \cdot \rangle$ denotes averaging in homogeneous directions to prevent numerical instabilities. The dynamic coefficient evaluation corresponds to minimizing the DMSRM modeling error
\begin{align*}
    \mathcal{E} = &\langle \mathcal{L}_{ij} \mathcal{L}_{ij} \rangle + (C^2_d)^2 \langle \mathcal{M}_{ij} \mathcal{M}_{ij} \rangle + \langle \mathcal{N}_{ij} \mathcal{N}_{ij} \rangle \\
    &- 2 ( C^2_d \langle \mathcal{L}_{ij} \mathcal{M}_{ij} \rangle + \langle \mathcal{L}_{ij} \mathcal{N}_{ij} \rangle - C^2_d \langle \mathcal{N}_{ij} \mathcal{M}_{ij} \rangle) \, .
\end{align*}
Finally, the DMSRM for the subfilter stress is obtained by filtering Eq.~\eqref{eqn:SRSSmodel_SRlevel} to the LES grid scale and downsampling  to $\overline{\Omega}$:
\begin{equation}
\tau^{\mathrm{DMSRM}}_{ij} =
    \underbrace{-2 C_{d}^2 (\Delta^*)^2 \check{\widehat{|{S}^*| S^*_{ij}}}}_{\text{DSM $\rightarrow$ DMSRM}}
    +
    \underbrace{(\check{\widehat{\widehat{{u}^*_i \, u^*_j}}} - \check{\widehat{\widehat{u^*_i} \, \widehat{u^*_j}}})}_{\text{SRM}}.
    \label{eqn:SRSSmodel_toUse}
\end{equation}
Here, the first term is the DSM contribution to the DMSRM, denoted ``DSM $\rightarrow$ DMSRM,'' and the second term denotes the \emph{super-resolution model} (SRM) contribution. 

It must be noted that in the DMSRM formulation, the DL component is employed exclusively to obtain $u_{i}^*$, from which the SFS stress tensor is computed using Eq.~\eqref{eqn:SRSSmodel_toUse}. This differs fundamentally from other DL-based LES closure approaches, which aim at predicting the SFS stress tensor directly from the resolved velocity field \citep{beck2019deep, gamahara2017searching}, optimize the neural network weights in conjunction with the governing flow equations to ensure \textit{a posteriori} consistency \citep{sirignano2020, macart_adjoint}, or derive a closure policy through reinforcement learning \citep{novati2021automating, kurz2023deep}.

 
\section{DNS datasets, preprocessing, and LES configurations}
\label{sec:dataset}

In this work, DNS data of forced HIT is first employed as it is a canonical configuration with well-characterized turbulence statistics and no boundary effects. These properties make it an established benchmark for evaluating turbulence closure models before extending them to more complex flows. 

We consider two forced HIT DNS datasets. The Taylor-microscale Reynolds numbers of the two DNSs are $\mathrm{Re_{\lambda}} \approx 110$ and $\mathrm{Re_{\lambda}} \approx 200$. These datasets are denoted as Re110 and Re200, respectively. Both simulations are performed on a triply periodic domain of size $L = 2\pi$ in each direction, which is uniformly discretized using $256^3$ points for Re110 and $512^3$ points for Re200. These grids satisfy $\mathrm{dx}/\eta \leq 2$, where $\mathrm{dx}$ is the uniform grid spacing and $\eta$ is the Kolmogorov length scale. The same configurations are also employed for \textit{a posteriori} LES analyses and verification, where the filter is implicitly defined by the LES grid, and the SFS and subgrid-scale stress tensor terminologies can be considered equivalent.

The simulations are performed using the CIAO code, which has been extensively applied in a variety of flow configurations~\citep{desjardins2008high, attili2012statistics, macart_adjoint, berger2025combustion}. The solver uses a staggered grid for the velocity components, with mass continuity enforced by solving a Poisson equation for the hydrodynamic pressure using preconditioned multigrid methods. The second-order Crank--Nicolson scheme is used for time advancement, and a fourth-order finite-difference scheme is applied for spatial discretization. Independent of whether the LES or DNS approach is considered, a sharp-spectral forcing with a cut-off wavenumber of $\kappa_{c} = 3$ is applied, similar to the setup employed by Eswaran \textit{et al.}~\cite{eswaran1988examination}. The DNS data are publicly available \citep{Nista_datapublication}, and their parameters are reported in table~\ref{tab:datasets}.

\begin{table}[!ht]
    \centering
    \begin{tabular}{c c c c c c c c}
    \toprule
    Dataset & $L$ & $N$ & $\mathrm{Re_{\lambda}}$ & $\mathrm{dx/\eta}$ & $L/\ell$ & Train & Test \\
    \midrule
    Re110 & $2\pi$ & $256^3$ & 110 & 1.58 & 5.26 & $\bullet$ &  \\
    Re200 & $2\pi$ & $512^3$ & 200 & 1.61 & 5.26 & $\bullet$ & $\bullet$ \\
    \bottomrule
    \end{tabular}
    \caption{Simulation parameters for each DNS dataset. ``$L$'' indicates the physical size of the periodic computational domain in each direction, \textit{N} is the number of mesh points, \textit{$Re_\lambda$} is the Taylor-microscale Reynolds number, \textit{$\mathrm{dx}/ \eta$} is the mesh resolution relative to the Kolmogorov microscale \textit{$\eta$}, and $L/\ell$ is the ratio of the domain size to the integral scale $\ell$. ``Train'' and ``Test'' flags indicate which data were used in SR-GAN training and \textit{a priori} and \textit{a posteriori} testing, respectively.}
    \label{tab:datasets}
\end{table}

To train the SR-GAN framework, 1400 snapshots of the three-dimensional velocity vector are collected from each of the Re110 and Re200 datasets. The snapshots are taken after an initial transient period of $t \approx 5 \tau_{0}$, where $\tau_{0}$ is the eddy-turnover time, with two snapshots extracted every $\tau_{0}$. Unlike conventional SR approaches that aim to directly reconstruct DNS-scale resolution, the SR-GAN framework in this work is trained to upsample to an intermediate resolution between filtered-DNS (F-DNS) fields, which are obtained using different filter widths. The input and target training pair consists of the following pairs of fields.
\begin{itemize}
    \item The low-resolution input fields, denoted  ``$\phi_{\mathrm{LR}}$,'' are obtained by filtering DNS snapshots with box, Gaussian, and spectral filter kernels using a filter width $\Delta_{\mathrm{LR}} = \mathrm{n}_{\Delta_\mathrm{{LR}}} \mathrm{dx}$. Here, $\mathrm{n}_{\Delta_{\mathrm{LR}}}$ specifies the LR filter size. Each F-DNS is then discretely downsampled by a factor of $\mathrm{n}_{\Delta_\mathrm{LR}}$ to produce the F-DNS field on a coarser grid, independent of the filter kernel applied. The definition of the filter kernels along with the downsampling operator is given in Appendix~\ref{sec:appendinx_filters}. The filter kernel to obtain $\phi_{\mathrm{LR}}$ is randomly selected during training to prevent the generator from learning an explicit inverse of the filters, as observed by Nista \textit{et al.}~\cite{nista2024PRF};
    
    \item The corresponding high-resolution target fields, denoted  ``$\phi_{\mathrm{HR}}$,''  are obtained from spectrally filtered DNS data using a smaller filter width $\Delta_{\mathrm{HR}} = \mathrm{n}_{\Delta_\mathrm{{HR}}} \mathrm{dx}$. Given the fixed upsampling ratio of the SR-GAN framework, the HR filter size is  $\mathrm{n}_{\Delta_{\mathrm{HR}}} = \mathrm{n}_{\Delta_{\mathrm{LR}}} / 2$.
\end{itemize}

Of the total number of extracted snapshots, 80$\%$ are used for the training dataset, while the remaining 20$\%$ are randomly divided between validation and testing. Four different training configurations are employed for the SR-GAN framework, and their performance is compared through \textit{a priori} and \textit{a posteriori} investigations.

\section{Generative adversarial framework and training strategies}
\label{sec:architecture}

\begin{figure}[!ht]
     \centering
     \includegraphics[width=\textwidth]{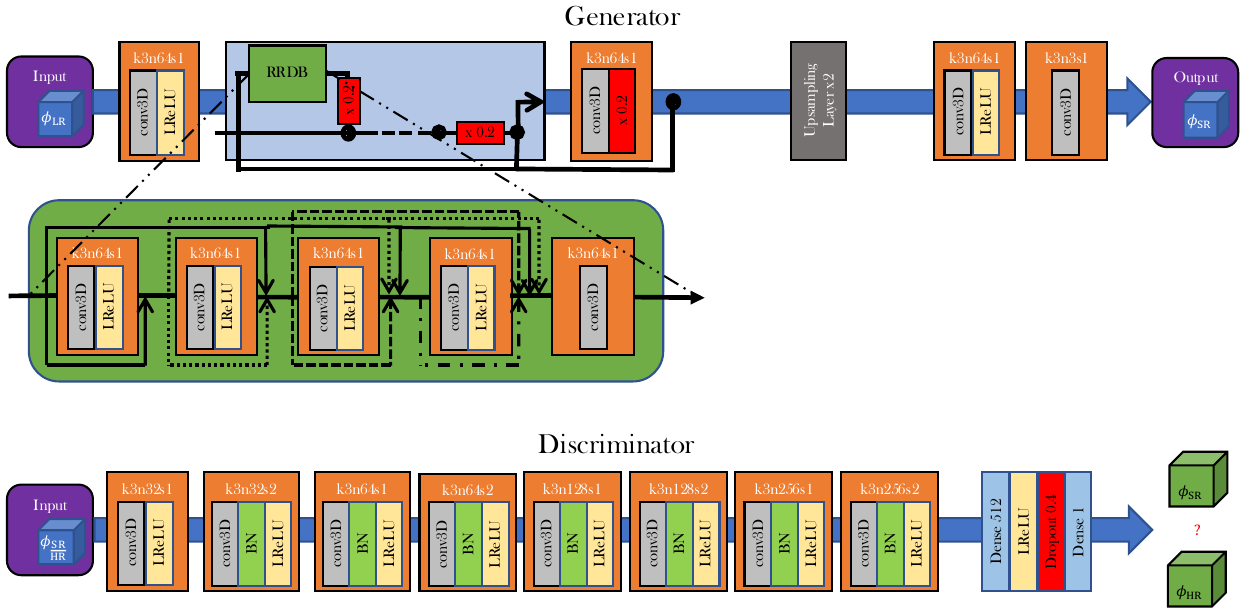}
     \caption{The generator (above) and discriminator (below) structures of the \emph{similarityGAN} framework, adapted from the ESRGAN architecture \citep{wang2018esrgan} and extended for three-dimensional turbulence SR reconstruction \citep{nista2024PRF}. In these illustrations, $\phi_\LRsub$ denotes the LR input field, $\phi_\mathrm{{SR}}$ denotes the SR output field, and $\phi_\mathrm{{HR}}$ denotes the corresponding HR target field. Each convolutional block contains kernels of size $k$, $n$ filter maps, and $s$ strides along each spatial dimension of the convolutional layer.}
     \label{fig:similarityGAN}
 \end{figure}

Figure~\ref{fig:similarityGAN} shows the \emph{similarityGAN} SR-GAN framework \citep{nista2024PRF, Nista2024_parallel} used in this work. \emph{SimilarityGAN} is designed for turbulence SR reconstruction and employs an adversarial training strategy using distinct generator and discriminator neural networks. This enables supervised, semi-supervised, and fully unsupervised learning without relying on user-defined objective functions \citep{goodfellow2020generative}. In GAN-based SR, the generator produces an SR version of $\phi_{\mathrm{LR}}$, denoted  $\phi_{\mathrm{SR}}$, while the discriminator network attempts to distinguish between $\phi_{\mathrm{SR}}$ and $\phi_{\mathrm{HR}}$. Throughout the training, the generator progressively learns to produce fields that are indistinguishable from $\phi_{\mathrm{HR}}$. In turn, the discriminator learns to judge the authenticity of the samples. It should be noted that the DMSRM formulation is agnostic to the specific SR framework employed. \emph{SimilarityGAN} is selected here due to its strong reconstruction performance on HIT flows compared to traditional fully supervised SR-based architectures \citep{nista2024PRF}.

The generator, shown in the upper row of figure~\ref{fig:similarityGAN}, is a CNN-based architecture that produces the $u^*_i$ field used by the DMSRM given the $\overline{u}_i$ input. Inspired by the work of Wang \textit{et al.}~\cite{wang2018esrgan}, the residual-in-residual dense block (RRDB) is used in this architecture, which comprises skip connections and three-dimensional convolution layers that allow for efficient feature learning and extraction of small-scale features. The upsampling operation is performed close to the network's output so that feature extraction and refinement occur at the  LR scale, which minimizes computational cost and memory requirements. As demonstrated by Ledig \textit{et al.}~\cite{ledig2017photo}, this design also ensures that the generator focuses on learning rich hierarchical representations before the spatial resolution is increased, therefore improving fidelity and training stability. The upsampling layer in the generator increases the spatial resolution by a factor of two in each dimension using nearest-neighbor interpolation, followed by a convolution operation to refine the output. The generator contains approximately 11 million trainable parameters. 

The discriminator, shown in the lower row of figure~\ref{fig:similarityGAN}, is structured as a deep convolutional network followed by fully connected layers, with leaky rectified linear unit (LReLU) activation functions similar to the generator but with a binary classification output. 
The discriminator has a depth comparable to that of the generator with approximately 15 million trainable parameters, ensuring balanced complexity during adversarial training. This discriminator structure facilitates efficient adversarial training and preserves its ability to discriminate between $\phi_{\mathrm{HR}}$ and $\phi_{\mathrm{SR}}$ \citep{nista2024PRF}.

\subsection{Training strategies}
\label{sec:PUT_training}

SR-GAN training proceeds in three stages: (1) independent, fully supervised training of the generator  (pretraining), (2) semi-supervised training of the SR-GAN, and (3) partially unsupervised training of the SR-GAN. These stages are explained below. The pretraining stage is discussed after the semi-supervised training stage.

During the semi-supervised training, the generator's loss function ($\mathcal{L}_{\mathrm{GEN}}$) consists of four contributions: a pixel-wise reconstruction loss ($L_{\mathrm{{pixel}}}$), a gradient-based loss to enforce small-scale fidelity ($L_\mathrm{{gradient}}$), a physics-based continuity loss ($L_\mathrm{{continuity}}$), and an adversarial loss ($L_\mathrm{{adversarial}}$), 
\begin{equation}
\begin{aligned}
     \mathcal{L}_{\mathrm{GEN}} &=  \alpha_1 \, L_{\mathrm{pixel}} + \alpha_2 \, L_\mathrm{{gradient}} + \alpha_3 \, L_\mathrm{{continuity}} + \alpha_4 \, L_\mathrm{{adversarial}} \\
     L_\mathrm{{pixel}} &= \mathrm{MSE}(\phi_\SRsub, \phi_\mathrm{HR}) \\
     L_\mathrm{{gradient}} &= \mathrm{MSE}(\grad \phi_\SRsub, \grad \phi_\mathrm{HR}) \\
     L_\mathrm{{continuity}} &= \mathrm{MSE}(\nabla \cdot \phi_\SRsub, 0) \\
     L_\mathrm{{adversarial}} &= - \mathbb{E}[\log(\sigma(D(G(\phi_\LRsub)) -\mathbb{E}[D(\phi_\mathrm{HR})]))] \\
     &\, \, \quad -\mathbb{E}[\log(1 - \sigma(D(\phi_\mathrm{HR}) - \mathbb{E}[D(G(\phi_\LRsub))]))] \, ,
\label{eqn:generator_loss}
\end{aligned}
\end{equation}
where $\mathbb{E}[\cdot]$ denotes the expectation over the training mini-batch, $\sigma(\cdot)$ is the sigmoid activation function, and $D$ and $G$ represent the discriminator's and generator's operators, respectively. The weighting coefficients are $\alpha_i = [0.88994, 0.06, 0.05, 6 \times 10^{-5}]$; these were chosen through hyperparameter tuning, such that the absolute value of each term in $\mathcal{L}_{\mathrm{GEN}}$ is of the same order during the semi-supervised training \citep{nista2024PRF}. The weighting of these coefficients is appropriate for the forced HIT datasets employed in this work, but may require adjustment for other flow configurations. Dynamic weighting strategies were not pursued due to the risk of destabilizing the adversarial training balance. All mean-squared error (MSE) terms are evaluated element-wise. The adversarial loss is based on the relativistic average GAN (RaGAN) definition proposed by Jolicoeur-Martineau~\cite{jolicoeur2018relativistic}, given its improved stability over the traditional adversarial loss. The discriminator is trained using the corresponding RaGAN-based loss, 
\begin{equation}
\begin{aligned}
\mathcal{L}_{\mathrm{DISC}} =\ &\mathbb{E}[\log(\sigma(D(\phi_\mathrm{HR}) - \mathbb{E}[D(G(\phi_\LRsub))]))]\ \\
&+ \mathbb{E}[\log(1 - \sigma(D(G(\phi_\LRsub)) - \mathbb{E}[D(\phi_\mathrm{HR})]))] \, .
\end{aligned}
\label{eqn:disc}
\end{equation}
The first term in Eq.~\eqref{eqn:disc} encourages the discriminator to assign higher fidelity scores to $\phi_\mathrm{HR}$ relative to the average score of $\phi_\mathrm{SR}$, while the second term penalizes the discriminator if the $\phi_\mathrm{SR}$ are rated more realistic than the average $\phi_\mathrm{HR}$.

Due to the well-known training instabilities associated with SR-GAN frameworks, an independent, fully supervised pretraining of the generator is employed before the semi-supervised SR-GAN training phase. In the pretraining phase, the generator is trained to minimize $L_{\mathrm{{pixel}}}$, $L_\mathrm{{gradient}}$, and $L_\mathrm{{continuity}}$. This means that the discriminator is not employed during this initial training phase. The generator is randomly initialized to avoid undue bias. This pretraining stage provides a well-conditioned and physically consistent starting point for the subsequent training phases. 

Recent studies have emphasized the critical role of the initial learning rate for local convergence and stability of SR-GAN training~\citep{GAN_lr}. Accordingly, learning rate choices for both the generator and discriminator follow recent recommendations shown to enhance GAN stability in turbulent flow contexts. The generator is pretrained with an initial learning rate of $1 \times 10^{-4}$, while the discriminator is initialized with a lower initial learning rate of $5 \times 10^{-5}$ to prevent its dominance over the generator, consistent with strategies reported by Mescheder \textit{et al.}~\cite{GAN_lr} and Nista \textit{et al.}~\cite{nista2021turbulent}. Both networks are optimized using the Adam algorithm with no learning rate schedule applied \citep{kingma2017adammethodstochasticoptimization}.

By training the discriminator to distinguish $\phi_\mathrm{HR}$ and $\phi_\mathrm{SR}$ during the semi-supervised training phase, physical information is implicitly embedded within its latent representation. As a result, the discriminator functions as a physics-informed loss component and guides the generator through adversarial feedback toward producing physically plausible $\phi_\mathrm{SR}$. This mechanism is further leveraged in a third training phase known as partially unsupervised training (PUT), introduced by Nista \textit{et al.}~\cite{nista2024PRF}. The PUT phase has been shown to be essential in improving the \textit{a priori} reconstruction and extrapolative capabilities of SR frameworks to unseen data, such as those generated with filter kernels different from those employed during training. In general, the broader the range of filter kernels and sizes employed during training, the more robust the SR reconstruction is expected to be in \textit{a posteriori} settings.

In the PUT phase, LES of forced HIT are performed using the same numerical solver described in Section~\ref{sec:dataset}, but on coarser grids consistent with the chosen $\mathrm{n}_{\Delta_\mathrm{LR}}$ used to generate the $\phi_{\mathrm{LR}}$. In these LES calculations, the unresolved scales are either closed using the DSM (Eq.~\eqref{eqn:DSM}), the DMM (Eq.~\eqref{eqn:DMM}), or their influence is not modeled (``implicitly modeled'' LES). These closure models are chosen due to their widespread adoption, though the framework is not limited to their use. 

When multiple $\mathrm{n}_{\Delta_\mathrm{LR}}$ are employed in the training setup, as is further described in Section~\ref{sec:training_setups}, LES calculations are performed at corresponding resolutions to provide matching input fields for the generator. The discriminator that has been trained during the semi-supervised phase is not further trained. Since no $\phi_\mathrm{HR}$ is available as a label during the PUT phase, the generator’s loss simplifies to its continuity and adversarial components,
\begin{equation}
\begin{aligned}
    \mathcal{L}_{\mathrm{GEN}_\mathrm{PUT}} &=  \alpha_3 \, L_\mathrm{{continuity}} + \alpha_4 \, L_\mathrm{{adversarial}_\mathrm{PUT}} \\
     L_\mathrm{{continuity}} &= \mathrm{MSE}(\nabla \cdot \phi_\SRsub, 0) \\
     L_\mathrm{{adversarial}_\mathrm{PUT}} &= - \mathbb{E}[\log(\sigma(D(G(\phi_\LRsub))))]\ ,
    \label{eqn:gen_loss_unsupervised}
\end{aligned}
\end{equation}
where the standard adversarial loss is employed. Thus, the PUT phase of the generator is exclusively driven by the discriminator’s physics-informed feedback and the continuity constraint.

\subsection{Training configurations}
\label{sec:training_setups}

Different training configurations are considered. They differ primarily in whether multiple $\phi_\mathrm{LR}$ of varying resolutions are used, denoted ``multiple,'' or a set of single fixed resolution is used, denoted ``single.'' In the ``single'' training configuration, the $\phi_\mathrm{LR}$ resolution is fixed at $\mathrm{n}_{\Delta_{\mathrm{LR}}}=16$ with the corresponding $\phi_\mathrm{HR}$ resolution at $\mathrm{n}_{\Delta_{\mathrm{HR}}}=8$. The model is trained exclusively for Re200 and incorporates the additional PUT phase. In the ``multiple'' training configuration, the model is trained simultaneously for $\mathrm{n}_{\Delta_{\mathrm{LR}}}\in\{16,14,12,10,8,6\}$  with a fixed upsampling factor of two. Thus, the corresponding $\phi_\mathrm{HR}$ counterparts are $\mathrm{n}_{\Delta_{\mathrm{HR}}}\in\{8,7,6,5,4,3\}$. This configuration is trained for both Re110 and Re200, and the additional PUT phase is optionally considered. Table~\ref{tab:training_configurations} summarizes the training configurations.
\begin{table}[!ht]
    \centering
    \begin{tabular}{c c c c c c c}
     \toprule
     \makecell{Training \\ configuration} & \makecell{Training \\ dataset} & $\mathrm{n_{\Delta_\mathrm{LR}}}$ & $\mathrm{n_{\Delta_\mathrm{HR}}}$ & $N_{\phi_{\mathrm{LR}}}$ & $N_{\phi_{\mathrm{HR}}}$ & \makecell{PUT} \\
     \midrule
     SG-Re110 (multiple) & Re110 & \makecell{16,14,12,\\ 10,8,6} & \makecell{8,7,6,\\5,4,3} & $16^3$ & $32^3$  & Yes \\
     \midrule
     SG-Re200 (single) & Re200 & 16 & 8 & $32^3$ & $64^3$ & Yes \\
     SG-Re200 (multiple) & Re200 & \makecell{16,14,12,\\10,8,6} & \makecell{8,7,6,\\5,4,3} & $32^3$ & $64^3$ & Yes \\
     SG-Re200 (multiple) - NoPUT & Re200 & \makecell{16,14,12,\\10,8,6} & \makecell{8,7,6,\\5,4,3} & $32^3$ & $64^3$ & No \\
     \bottomrule
    \end{tabular}
    \caption{Summary of the training configurations used for the \emph{similarityGAN} (SG) framework. $\mathrm{n_{\Delta_\mathrm{LR}}}$ and $\mathrm{n_{\Delta_\mathrm{HR}}}$ denote the filter sizes applied to generate $\phi_{\mathrm{LR}}$ and $\phi_{\mathrm{HR}}$ from the DNS and employed during the training. $N_{\phi_{\mathrm{LR}}}$ and $N_{\phi_{\mathrm{HR}}}$ specify the sizes of the corresponding training subbox pairs. The column ``PUT'' indicates whether the partially unsupervised training phase is applied.
    }
    \label{tab:training_configurations}
\end{table}

Due to the large amount of training data, loading the entire dataset into GPU memory is impractical. To address this, a patch-to-patch training strategy is applied, in which subboxes are randomly extracted from the full domain for each snapshot \citep{Nista2024_parallel}. For simplicity, the same training subbox sizes are used across all filter sizes, namely $64^3$ for $\phi_{\mathrm{HR}}$ and $32^3$ for $\phi_{\mathrm{LR}}$ for the Re200 dataset, and $32^3$ and $16^3$, respectively, for the Re110 dataset, with an equal number of subboxes per filter regardless of $\mathrm{n}_{\Delta_{\mathrm{LR}}}$. These training subbox sizes are found to be suitable for training due to the batch size and box size trade-off associated with finite memory capacity \citep{Nista2024_parallel}. 

All training is performed on the CLAIX cluster at RWTH Aachen University using a single node equipped with four NVIDIA H100 GPUs (96 GB each). The total training time of the SG-Re200 (single) setup is approximately 4 hours. When training simultaneously using multiple filter sizes -- SG-Re200 (multiple) -- the training time increases to approximately 42 hours. The most computationally demanding training configuration is the one with the PUT phase enabled, for which the training time increases to approximately 65 hours. The computational time required for the LES SR predictions themselves, as well as the associated preprocessing steps, further increases the total time but is not considered in the values mentioned above. For the SG-Re110 (multiple) setup, the training time is reduced by approximately 30\% owing to the smaller training subbox pairs.


\section{\emph{A priori} investigations of the DMSRM on forced HIT}
\label{sec:apriori_sec}

\begin{figure}[!ht]
  \centering
  \begin{minipage}[b]{0.495\textwidth}
  \includegraphics[width=\textwidth]{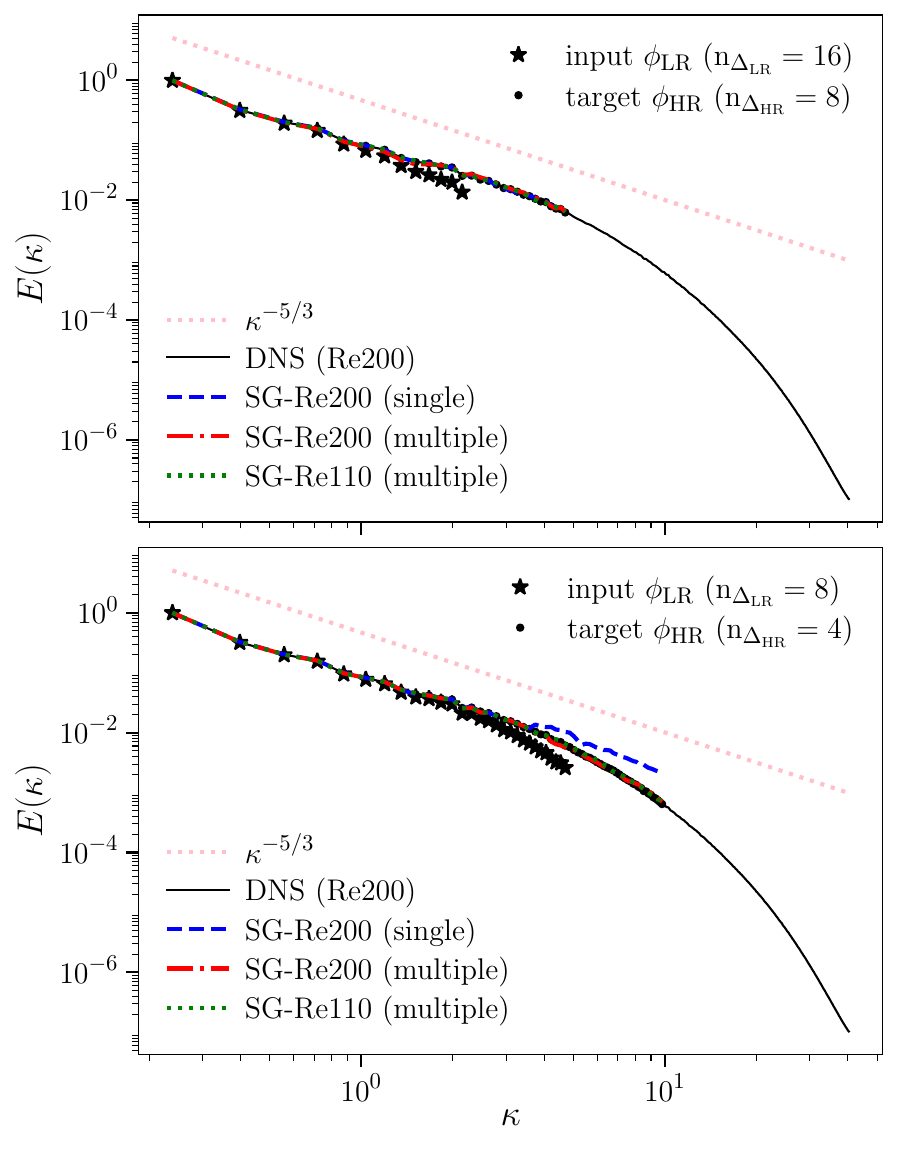}
  \end{minipage}
  \hfill
  \begin{minipage}[b]{0.495\textwidth}
  \centering
  \includegraphics[width=\textwidth]{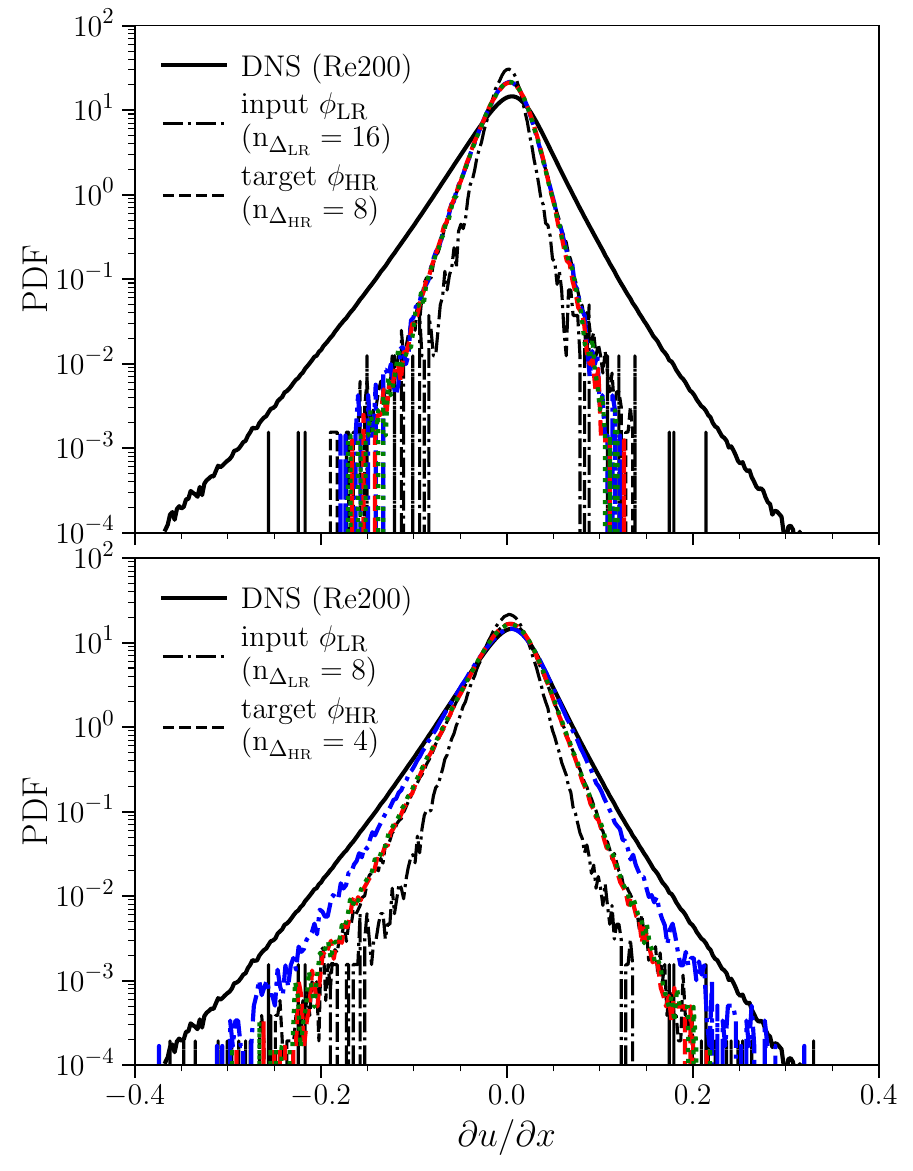}
  \end{minipage}
  \caption{\textit{A priori} normalized TKE spectra (left) and PDFs of $\partial u / \partial x$ (right) for ``out-of-sample'' input $\phi_{\mathrm{LR}}$ fields from the Re200 dataset. The top row shows results for $\mathrm{n}_{\Delta_\mathrm{LR}}=16$, and the bottom row for $\mathrm{n}_{\Delta_\mathrm{LR}}=8$. ``SG-Re200 (single)'' and ``SG-Re200 (multiple)'' refer to the \emph{similarityGAN} framework trained on the Re200 dataset using either a single fixed $\phi_{\mathrm{LR}}$ resolution or multiple varying $\phi_{\mathrm{LR}}$ resolutions, respectively. ``SG-Re110 (multiple)'' refers to the same framework trained on the Re110 dataset with multiple varying $\phi_{\mathrm{LR}}$ resolutions (see table~\ref{tab:training_configurations}). For clarity, red and green lines overlap with the target $\mathrm{\phi_{HR}}$ line.}
  \label{fig:TKE_PDF_apriori}
\end{figure}

We first investigate the reconstruction capabilities of the \emph{similarityGAN} framework. Specifically, given $\phi_{\mathrm{LR}}$ at resolution $\mathrm{n}_{\Delta_\mathrm{LR}}$, \emph{similarityGAN} generates an SR field at resolution $\mathrm{n}_{\Delta_\mathrm{HR}}$, aiming to generate the corresponding $\phi_{\mathrm{HR}}$. In this analysis, $\phi_{\mathrm{LR}}$ is obtained using a filter kernel different from those employed to generate the training $\phi_{\mathrm{LR}}$. This ``out-of-sample'' kernel distorts the turbulent kinetic energy (TKE) distribution at scales larger than the filter cut-off, mimicking the numerical errors typically encountered in LES. Additional details are given in Appendix~\ref{sec:appendinx_filters}. Since  \emph{similarityGAN}'s reconstruction performance for $\phi_{\mathrm{LR}}$ obtained using ``in-sample'' filter kernels is known to be strong \citep{nista2024PRF}, we do not repeat the ``in-sample'' analysis here. We conduct subsequent analyses using the Re200 dataset, which allows for large filter widths within the inertial subrange, making the results more relevant to LES of higher Reynolds number flows.

\begin{figure}[!ht]
    \centering
    \includegraphics[width=\textwidth]{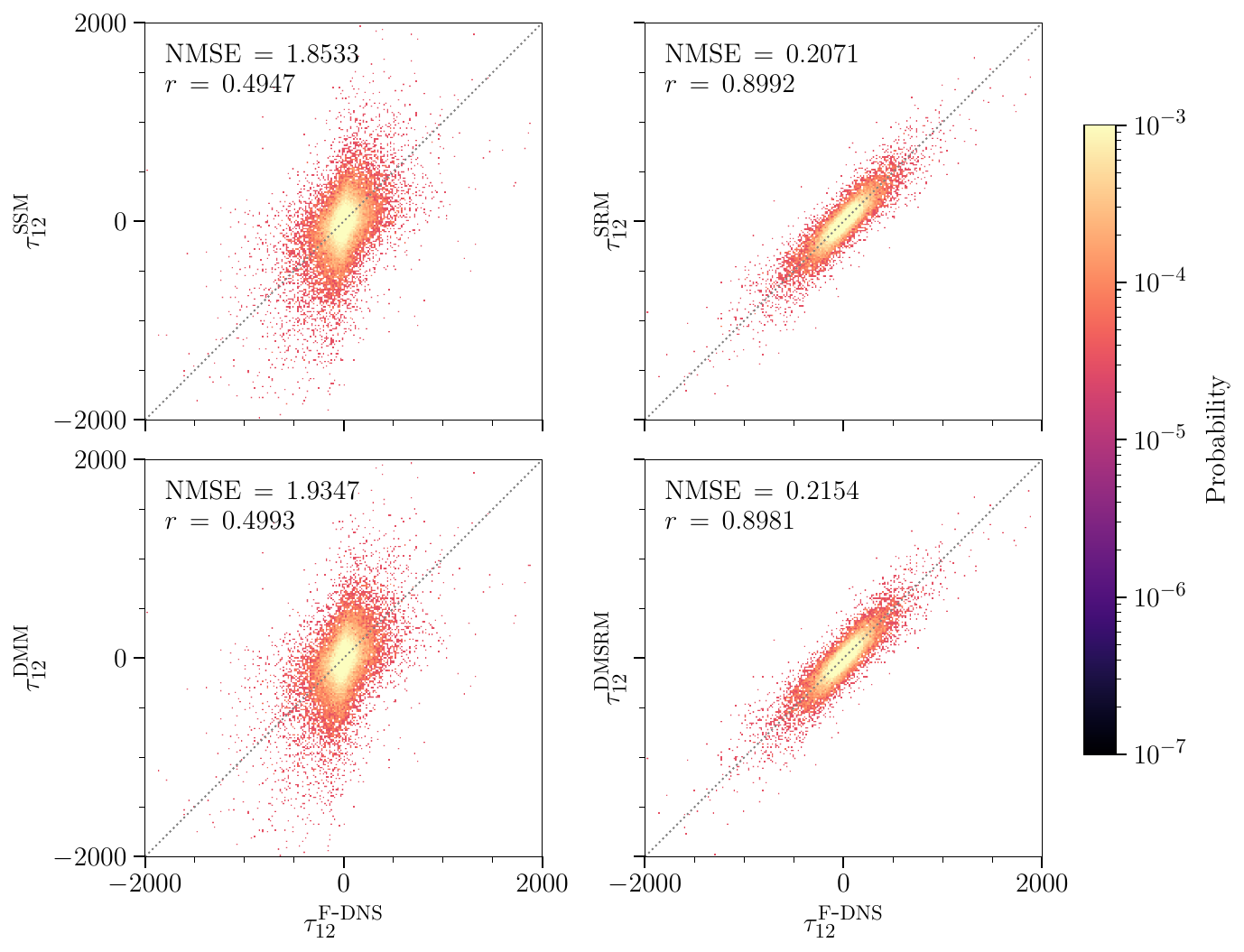}
    \caption{jPDFs of the F-DNS reference of the SFS stress tensor component $\tau^{\mathrm{F\mbox{-}DNS}}_{12}$ and the SSM component $\tau^{\mathrm{SSM}}_{12}$ (upper left), the SRM component $\tau^{\mathrm{SRM}}_{12}$ (upper right), the  DMM component $\tau^{\mathrm{DMM}}_{12}$ (lower left), and the DMSRM component $\tau^{\mathrm{DMSRM}}_{12}$ (lower right), evaluated for a test snapshot of the Re200 dataset at $\mathrm{n}_{\Delta_\mathrm{LR}} = 16$. SRM and DMSRM $\tau_{12}$ components are evaluated using the SR field obtained using the SG-Re200 (multiple) training configuration. ``NMSE'' indicates the MSE normalized by the variance of the F-DNS reference field $\tau^{\mathrm{F\mbox{-}DNS}}_{12}$, evaluated over the spatial domain of the test snapshots.}
    \label{fig:jPDF_apriori_Delta16}
\end{figure}

Figure~\ref{fig:TKE_PDF_apriori} shows TKE spectra (left) and probability density functions (PDFs) of one component of the velocity gradient, $\partial u / \partial x$, (right) of the SR fields obtained using the three training configurations with the PUT phase, as described in table~\ref{tab:training_configurations}. Results are given for two input $\phi_\mathrm{LR}$ resolutions: $\mathrm{n}_{\Delta_\mathrm{LR}}=16$ in the top row and $\mathrm{n}_{\Delta_\mathrm{LR}}=8$ in the bottom row. The SG-Re200 (multiple) training configuration demonstrates strong reconstruction performance at both resolutions. The TKE spectra closely match the target $\phi_\mathrm{HR}$, and PDFs of $\partial u / \partial x$ reflect accurate fine-scale statistics, even under ``out-of-sample'' filter kernel conditions. In contrast, the SG-Re200 (single) training configuration performs well only for $\mathrm{n}_{\Delta_\mathrm{LR}}=16$, but fails to reproduce small-scale velocity fluctuations and overpredicts the TKE spectra for the unseen input resolution $\mathrm{n}_{\Delta_\mathrm{LR}}=8$. These results indicate that training with $\phi_\mathrm{LR}$ of varying resolutions preserves the reconstruction accuracy across different input resolutions, relaxing the requirement of matching the $\Delta/\eta$ ratio between training and testing conditions~\citep{nista2024PRF}. This is particularly important for \textit{a posteriori} deployment of the \emph{similarityGAN} framework, even when training and testing are performed using the same dataset, since the LES filter kernel and size are not known \citep{arumapperuma2025extrapolation}.

Another important aspect for practical \textit{a posteriori} deployment is generalizability to higher Reynolds numbers. This is demonstrated in figure~\ref{fig:TKE_PDF_apriori}, which shows TKE spectra and velocity-gradient PDFs of SR fields obtained by applying the SG-Re110 (multiple) model to upsample Re200 $\phi_\mathrm{LR}$ fields at resolutions of $\mathrm{n}_{\Delta_{\mathrm{LR}}}=16$ and $\mathrm{n}_{\Delta_{\mathrm{LR}}}=8$. Despite being trained for lower Reynolds number flow, the SG-Re110 (multiple) training configuration exhibits good performance when applied to the Re200 dataset, closely reconstructing the TKE spectra and velocity-gradient PDF at the two $\mathrm{n}_{\Delta_{\mathrm{LR}}}$ ratios. We therefore select the ``multiple'' training configuration for the DMSRM (Eq.~\eqref{eqn:SRSSmodel_toUse}) for all subsequent analyses.

The effectiveness of DMSRM is further examined through \textit{a priori} correlations of the SFS stress tensor component $\tau_{12}$.  Figure~\ref{fig:jPDF_apriori_Delta16} shows joint PDFs (jPDFs) of the FDNS-evaluated SFS stress component $\tau_{12}$ and the analogous component evaluated using the DMM (left) and the DMSRM (right). The top row of figure~\ref{fig:jPDF_apriori_Delta16} includes only the scale-similarity and super-resolution contributions; the bottom row includes the dynamic Smagorinsky contribution. The Pearson correlation coefficient, $r$, for the SRM is approximately twice that of the SSM, indicating a significantly improved agreement with the DNS. The superior performance of the SRM demonstrates that estimating the scale-similarity term from the SR velocity field provides more accurate SFS stress predictions than the standard SSM.

It is worth noting that the distributions obtained with the DMM and DMSRM closely match the corresponding scale-similarity and super-resolution contributions, as also reflected by similar $r$ values, indicating that the scale-similarity term is much larger than the DSM contribution. Interestingly, the dynamic Smagorinsky contribution to the DMM marginally increases the Pearson correlation coefficient, while in the DMSRM, it marginally decreases it. While not shown here, similar trends are observed for the finer input resolution with $\mathrm{n}_{\Delta_\mathrm{LR}} = 8$. 

It is important to note that the Pearson correlation coefficient for the SSM is approximately 30$\%$ lower than values typically reported in other \textit{a priori} studies, e.g., \cite{bardina1983improved, meneveau2000scale, liu1994properties}, which are often above 0.7. This difference is related to the specific approach used to compute the input velocity field $\overline{u}_{i}$ in \textit{a priori} settings, as detailed in Appendix~\ref{sec:correlation_SS}. 

\begin{figure}[!ht]
  \centering
  \begin{minipage}[b]{0.495\textwidth}
  \centering
  \includegraphics[width=\textwidth]{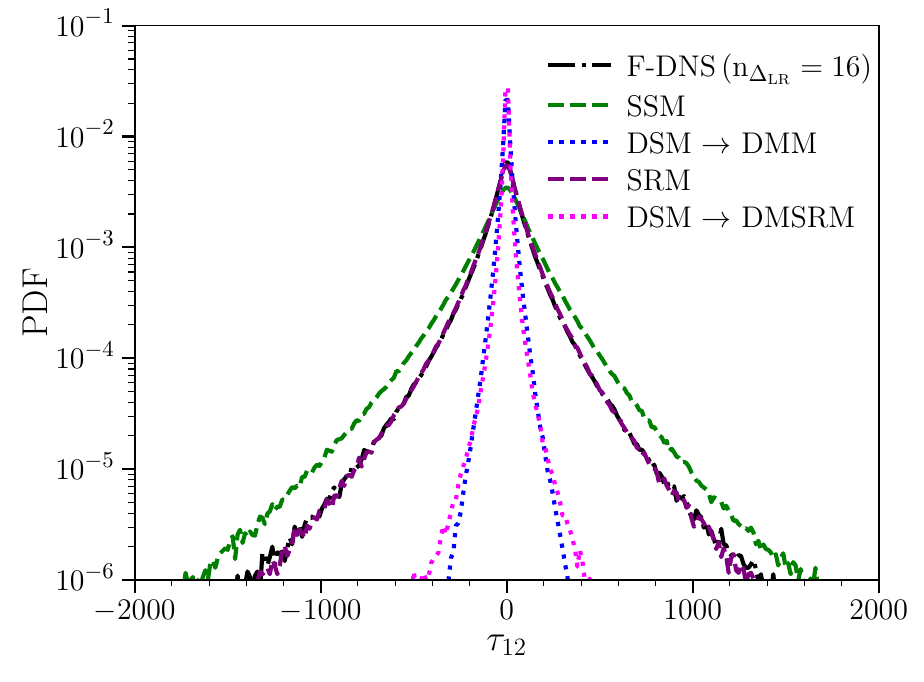}
  \end{minipage}
  \hfill
  \begin{minipage}[b]{0.495\textwidth}
  \centering
  \includegraphics[width=\textwidth]{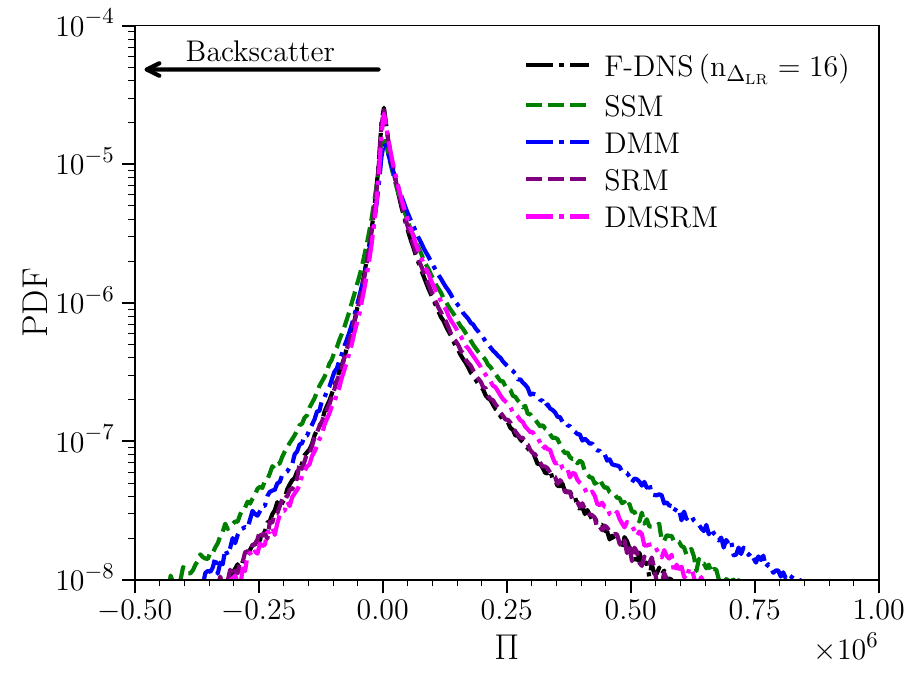}
  \end{minipage}
  \caption{PDFs of $\tau_{12}$ (left) and $\Pi$ (right) obtained from the DMM, DMSRM, and their individual components compared against the F-DNS reference (see Section~\ref{sec:dynamic_mixed_SR_SS_model}), using the Re200 dataset. } 
\label{fig:apriori_tau_12_energy_flux_contributions}
\end{figure}

Further insight is provided in figure~\ref{fig:apriori_tau_12_energy_flux_contributions} (left), which shows the PDFs of $\tau_{12}$ obtained from the individual components of the DMM and DMSRM (see Eqs.~\eqref{eqn:DMM} and \eqref{eqn:SRSSmodel_toUse}) compared against the F-DNS reference. The PDF of the SRM nearly overlaps with the F-DNS, both in peak location and tail behavior, suggesting that the SRM accurately reconstructs the SFS stress both pointwise (figure~\ref{fig:jPDF_apriori_Delta16}, top-right) and statistically. In contrast, the PDF of the SSM overpredicts high-magnitude stresses, characterized by a broader distribution compared to the F-DNS, which is consistent with prior observations (figure~\ref{fig:jPDF_apriori_Delta16}, top-left). As is also evident from the previous analysis, the scale-similarity components clearly dominate the PDF of $\tau_{12}$ in the dynamic mixed formulations, as both the ``DSM $\rightarrow$ DMM'' and ``DSM $\rightarrow$ DMSRM'' PDFs are significantly narrower.

The SFS energy dissipation, $\Pi = - \tau_{ij} \overline{S}_{ij}$, quantifies the energy transfer between resolved and subfilter scales, with positive values indicating dissipation and negative values indicating backscatter. Figure~\ref{fig:apriori_tau_12_energy_flux_contributions} (right) also shows the PDFs of $\Pi$ obtained with the DMM, DMSRM, and their individual components compared against the F-DNS. As expected, the SSM exhibits excessive backscatter, a behavior that can lead to numerical instabilities in \textit{a posteriori} simulations. The SSM also overpredicts SFS energy dissipation. Combining the SSM with the DSM (i.e., the DMM) mitigates the excessive backscatter thanks to the DSM's dissipative nature. However, this correction comes at the cost of increased dissipation relative to the F-DNS. Conversely, the SRM alone provides a much closer match to the F-DNS in terms of dissipation and backscatter. When combined with the DSM, a slight overprediction of the dissipation is apparent, while its impact on backscatter remains minimal. Interestingly, while the DMM yields an improved prediction of SFS energy dissipation compared to the SSM, it still overpredicts the stress tensor components. In contrast, the SRM alone already achieves an accurate reconstruction of the SFS stress components and the SFS energy dissipation. This \textit{a priori} analysis suggests that the inclusion of the DSM may introduce excessive dissipation, potentially degrading the DMSRM’s overall fidelity, though it may still be required for numerical stability in \textit{a posteriori} settings, which we investigate in the next section.


\section{\emph{A posteriori} investigations of the DMSRM on forced HIT}
\label{sec:aposteriori_sec}

\subsection{Physical accuracy of the DMSRM}

\begin{figure}[!ht]
  \centering
  \begin{minipage}[b]{0.495\textwidth}
  \includegraphics[width=\textwidth]{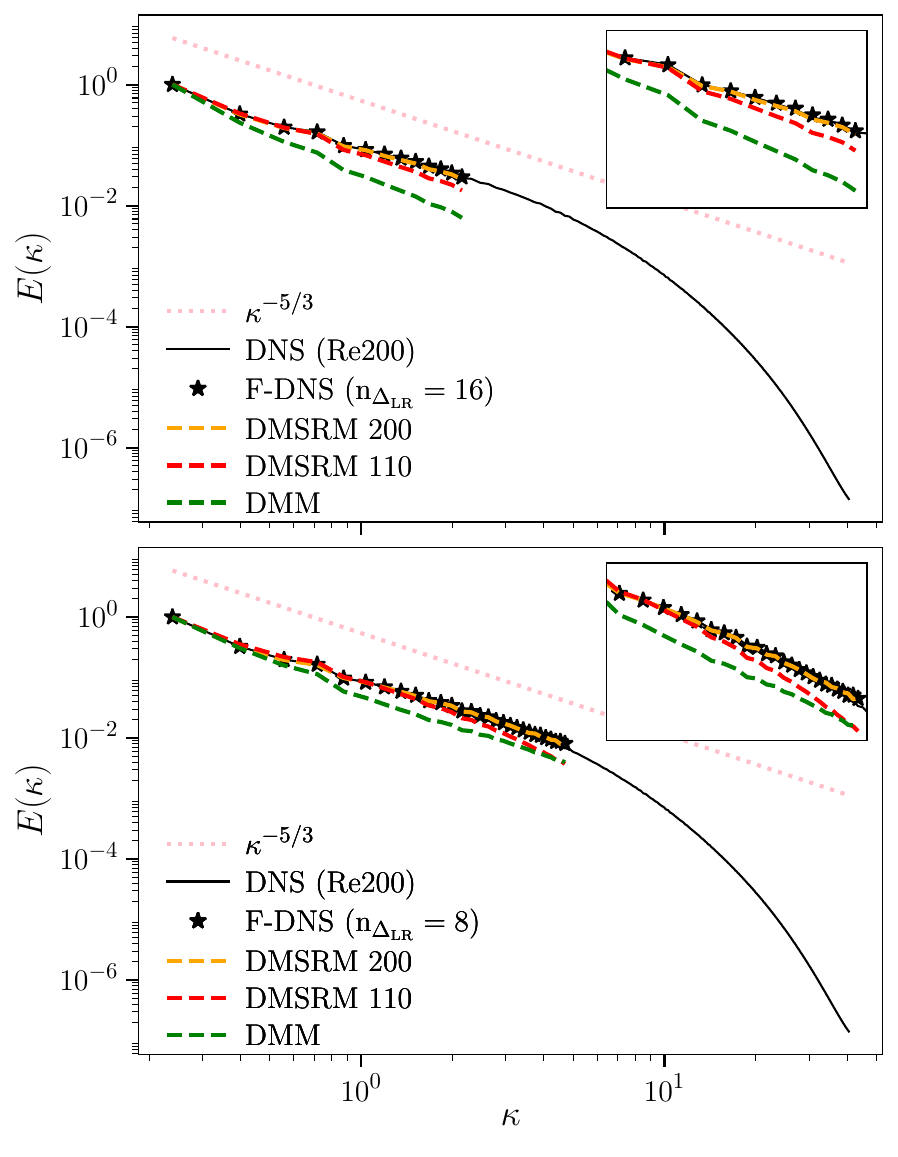}
  \end{minipage}
  \hfill
  \begin{minipage}[b]{0.495\textwidth}
  \centering
  \includegraphics[width=\textwidth]{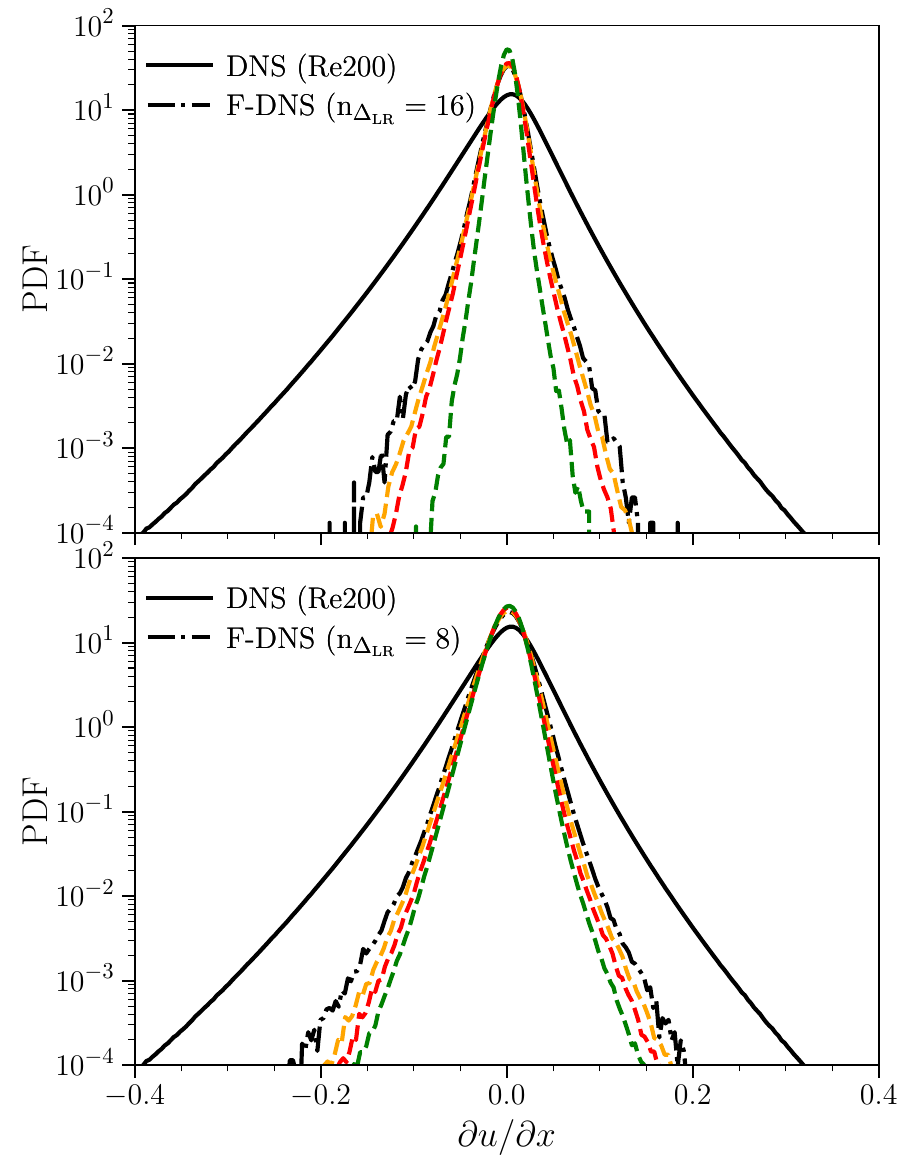}
  \end{minipage}
  \caption{\textit{A posteriori} normalized TKE spectra (left) and PDFs of $\partial u/\partial x$ (right) of LES coarsened by $\mathrm{n}_{\Delta_\mathrm{LR}}=16$ (top row) and $\mathrm{n}_{\Delta_\mathrm{LR}}=8$ (bottom row) from the Re200 DNS.  F-DNS fields are obtained with a spectral filter kernel operator with consistent ${\Delta_\mathrm{LR}}$. The ``DMSRM 200'' uses the SG-Re200 (multiple) model, while ``DMSRM 110'' uses the SG-Re110 (multiple).}
  \label{fig:TKE_PDF_aposteriori}
\end{figure}

We next evaluate the predictive fidelity of the DMSRM using \textit{a posteriori} LES. We simulate the Re200 configuration (see Section~\ref{sec:dataset}) using the same forcing and two LES grids of resolution $32^3$, corresponding to $\mathrm{n_{\Delta_{\mathrm{LR}}}} = 16$ ($\Delta_{\mathrm{LR}}/\eta \approx 26$), and $64^3$, corresponding to $\mathrm{n_{\Delta_{\mathrm{LR}}}} = 8$ ($\Delta_{\mathrm{LR}}/\eta \approx 13$). Two DMSRM configurations are considered. ``DMSRM 200'' employs the \emph{similarityGAN} model trained on the SG-Re200 (multiple) training,  while ``DMSRM 110''  uses the SG-Re110 (multiple) model. The latter is used to evaluate the DMSRM’s ability to generalize to a higher Reynolds number flow. Results are compared against the reference F-DNS, obtained with a spectral filter kernel operator with consistent $\mathrm{{\Delta_{\mathrm{LR}}}}$.

Figure~\ref{fig:TKE_PDF_aposteriori} shows the TKE spectra (left) and PDFs of $\partial u/ \partial x$ (right) for the coarse (top row) and fine (bottom row) LES resolutions. At the coarse resolution ($\mathrm{n}_{\Delta_{\mathrm{LR}}} = 16$), both DMSRMs are in close agreement with the F-DNS and outperform the DMM. The differences between the two DMSRM configurations are small. DMSRM 200 achieves the most accurate performance, captures the inertial subrange scaling, and maintains TKE levels close to those of the F-DNS throughout all turbulent scales. Good agreement is also observed for the velocity gradient PDFs, which show well-predicted heavy tails, indicating successful modeling of small-scale intermittency. DMSRM 110 shows similar trends, albeit with a small underprediction of TKE near the grid cut-off (shown in the insets) and a corresponding narrowing of the velocity gradient PDF. This behavior is linked to the slightly degraded SR reconstruction obtained by SG-Re110 on Re200 LES inputs compared to that of SG-Re200, as further demonstrated in Section~\ref{sec:aposteriori_stability}. The DMM exhibits a clear deviation from the TKE spectra, particularly by underpredicting intermediate- and high-wavenumber energy, which reflects its overly dissipative behavior. This is also seen in its narrower velocity gradient PDF, indicating suppressed turbulence intermittency. At the finer resolution ($\mathrm{n}_{\Delta_{\mathrm{LR}}} = 8$), the differences between all dynamic mixed models decrease, while their overall trends remain similar. It is worth noting that DMSRM 200 at $\mathrm{n_{\Delta_\mathrm{LR}}} = 16$ achieves improved accuracy relative to DMM at $\mathrm{n_{\Delta_\mathrm{LR}}} = 8$, i.e., DMM executed on a grid with the same resolution as ${\Omega}^{*}$, in terms of both TKE spectra and velocity gradient PDFs. This comparison is particularly relevant since the DMSRM 200 at $\mathrm{n_{\Delta_\mathrm{LR}}} = 16$ employs ${\Omega}^{*}$ only for SR inference and SFS stress computation, while solving the LES governing equations exclusively on $\overline{\Omega}$.

Moreover, the robustness of the DMSRM to the numerical discretization order is further assessed in Appendix~\ref{sec:appendix_discretization}, where additional \textit{a posteriori} LES with a second-order finite-difference scheme are reported. Similarly to the results reported above, the DMSRM 200 with second-order discretization still outperforms the DMM with fourth-order discretization, with only a small TKE underprediction near the grid cut-off and a corresponding narrowing of the velocity gradient PDF, which is slightly more pronounced at the coarser resolution.

\begin{figure}[!ht]
  \centering
  \begin{minipage}[b]{0.495\textwidth}
  \includegraphics[width=\textwidth]{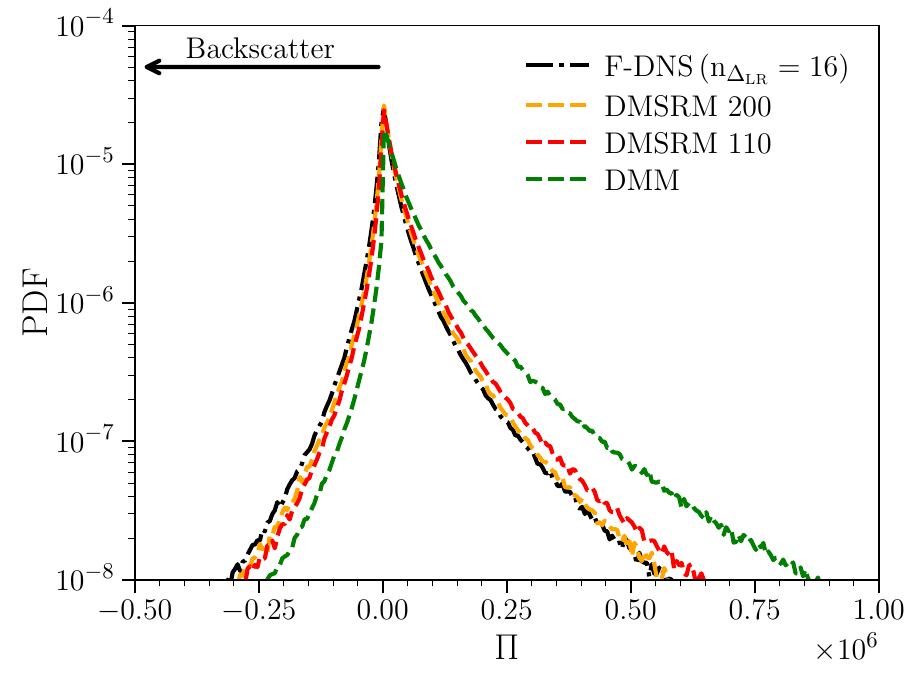}
  \end{minipage}
  \hfill
  \begin{minipage}[b]{0.495\textwidth}
  \centering
  \includegraphics[width=\textwidth]{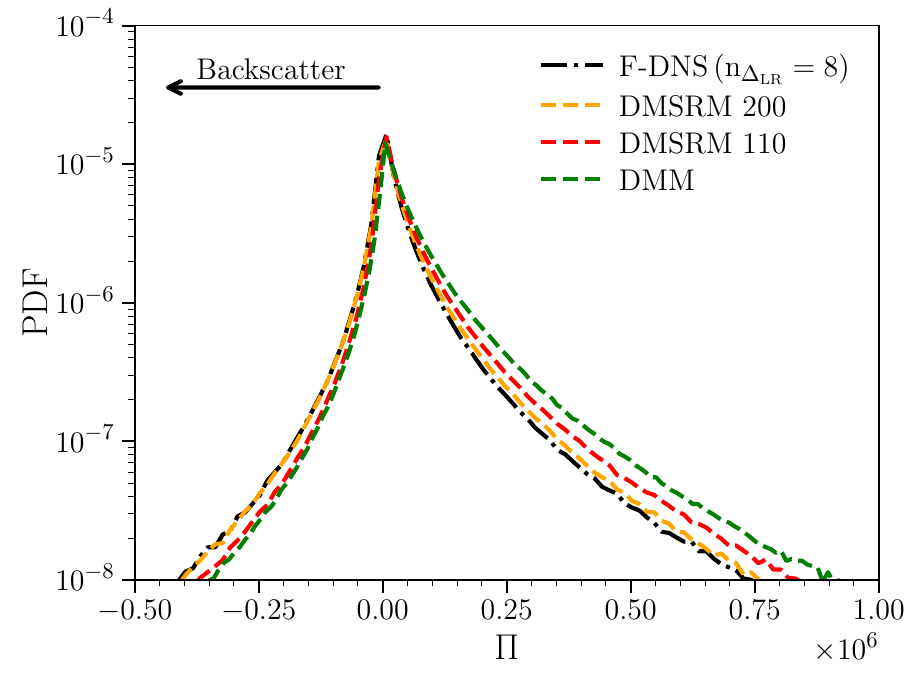}
  \end{minipage}
  \caption{\textit{A posteriori} PDFs of $\Pi$ for both DMSRM setups and the DMM against the F-DNS reference at coarser (left) and finer (right) resolutions.}
  \label{fig:SGSdissipation_aposteriori}
\end{figure}

To provide insight into the reasons for the above observations, figure~\ref{fig:SGSdissipation_aposteriori} shows PDFs of $\Pi$ for both DMSRMs and the DMM against the F-DNS reference at the coarse (left) and fine (right) resolutions. At the coarse resolution, both DMSRMs exhibit good agreement with the F-DNS, with DMSRM 200 achieving the closest agreement with the F-DNS across both positive and negative values of $\Pi$, while DMSRM 110 is slightly more dissipative. In contrast, the DMM substantially overpredicts the SFS energy dissipation and underestimates backscatter. This imbalance is rooted in a well-known limitation of SSMs, which tend to generate excessive backscatter and can become unstable due to uncontrolled energy transfer from the subfilter to the resolved scales. To mitigate this effect, DMMs introduce a dissipative DSM term, but this correction leads to overly damped turbulence dynamics and suppression of backscatter. This observation is consistent with the \textit{a priori} analyses presented in Section~\ref{sec:apriori_sec}.

These observations are quantified through the relative impact of backscatter to the total SFS energy dissipation, which is defined as
\begin{equation}
B_{\Pi} = - \frac{\langle \min(\Pi^{\mathrm{model}},0) \rangle_t}{\langle \Pi^{\mathrm{model}} \rangle_t} \, \mathrm{with} \, \Pi^{\mathrm{model}} = - \tau^{\mathrm{model}}_{ij} \overline{S}_{ij} \, ,
\label{eqn:backscatter}
\end{equation}
where $\langle \cdot \rangle_t$ denotes averaging over the spatial domain and an ensemble of LES snapshots, and the superscript denotes the selected closure model for $\tau_{ij}$. At the coarse resolution, backscatter computed from F-DNS accounts for approximately 20$\%$ of $\Pi$. The SRM contributions obtained with SG-Re200 and SG-Re110 yield backscatter levels of approximately 19$\%$ and 18$\%$, respectively, close to the F-DNS. When these SRMs are combined with the DSM contribution, the DMSRM net backscatter remains similar, with values of approximately 18$\%$ (SG-Re200) and 16$\%$ (SG-Re110). In contrast, the SSM alone predicts a larger fraction of 27$\%$, which is reduced substantially to approximately 9$\%$ with the DMM formulation. While this reduction enhances numerical stability, it also reflects the overly dissipative nature of the DMM. At the fine resolution, the overprediction of SFS energy dissipation for the DMM is decreased, and backscatter increases (to approximately $11\%$) and approaches the F-DNS value (approximately $13\%$). Both DMSRMs maintain their accurate backscatter prediction, with backscatter values very close to the F-DNS value. 

This trend is also reflected in the averaged $C_d^2$, which is evaluated from the DSM contributions to the DMSRM and DMM. At the coarse resolution, values of 0.07 and 0.09 are obtained for DMSRM 200 and DMSRM 110, respectively, compared to 0.13 for the DMM. A similar pattern with decreased values is observed at the fine resolution. Additionally, the DMM exhibits greater $C_d^2$ scatter than the DMSRMs, which is consistent with its less accurate representation of local SFS stresses \citep{liu1995experimental}.

These results highlight that an accurate scale-similarity contribution is essential for achieving close agreement with the F-DNS. Backscatter is reproduced through the scale-similarity term, which in turn controls the DSM contribution and ensures consistent mean energy transfer across scales. In this regard, the DMSRM outperforms the DMM. Its advantage lies in the data-driven SR-GAN framework, which is trained on high-fidelity data and optimized with an adversarial loss; therefore, physically plausible super-resolved velocity fields are reconstructed. The SFS stress tensor at the grid-filter scale is then obtained directly from such SR fields using a filter kernel and size consistent with training, yielding SFS stresses of the correct magnitude that accurately reproduce the dissipation. This reduces reliance on additional dissipation provided by the DSM term, which is smaller compared to the DMM, and thus maintains physically consistent backscatter. As a result, the SR-based approach produces a more physically and numerically consistent reconstruction than both the LES and SR test-filtered fields, making it a robust alternative to the traditional SSM in the DMM formulation.

\subsection{Stability of the DMSRM}
\label{sec:aposteriori_stability}

The previous analyses of the DMSRM show that the DSM contribution is small, raising the question of its necessity within the model formulation. In the absence of the DSM contribution, the DMSRM reduces to the SRM (Eq.~\eqref{eqn:SRSSmodel_toUse}). For the following tests, the SG-Re200 (multiple) training is employed, and the coarse resolution with $\mathrm{n}_{\Delta_\mathrm{LR}}=16$ is considered. This variant is hereafter referred to as ``SRM 200''. 

\begin{figure}[!ht]
  \centering
  \begin{minipage}[b]{0.495\textwidth}
  \includegraphics[width=\textwidth]{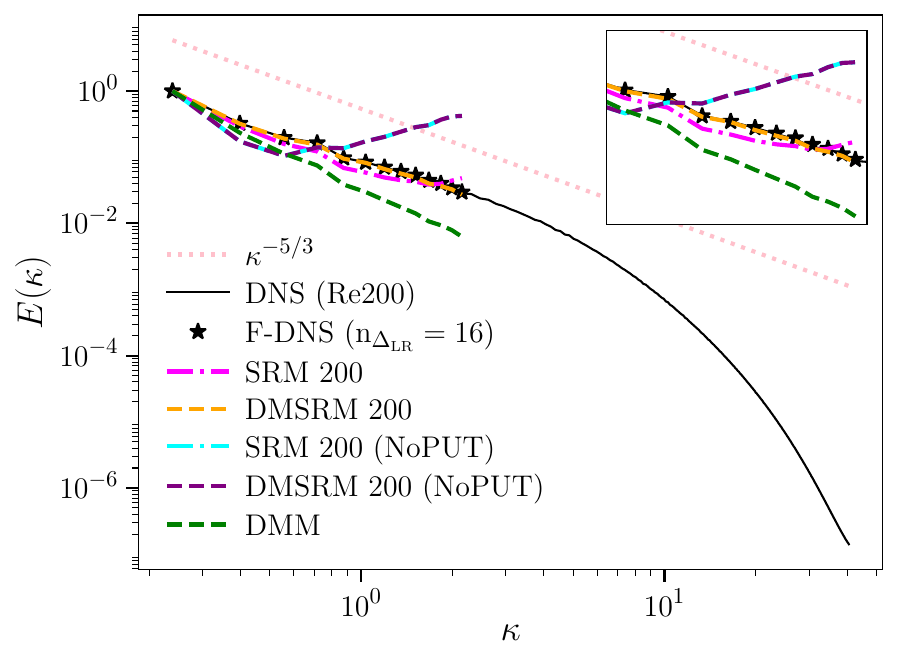}
  \end{minipage}
  \hfill
  \begin{minipage}[b]{0.495\textwidth}
  \centering
  \includegraphics[width=\textwidth]{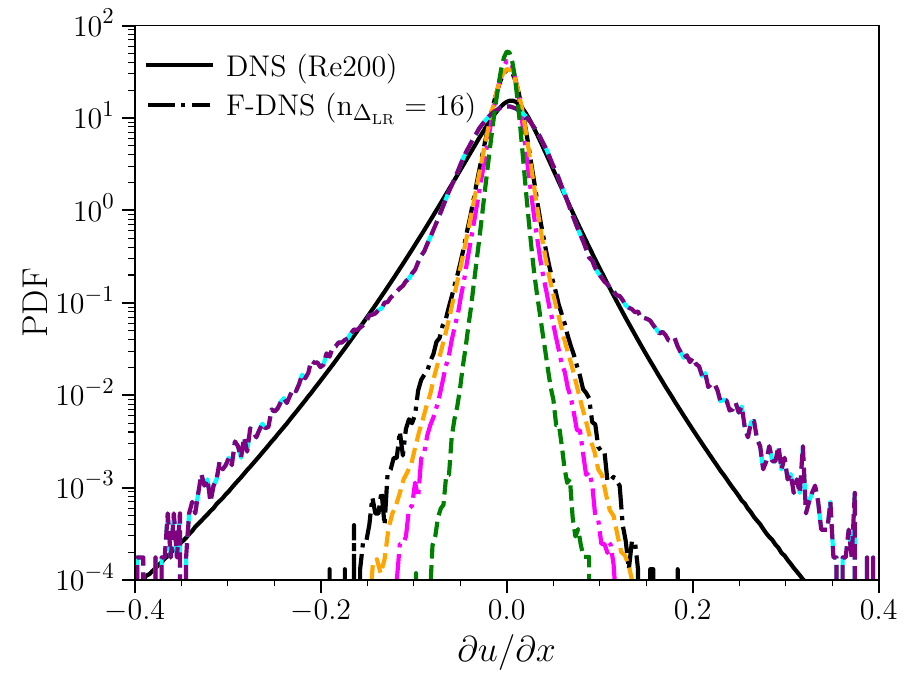}
  \end{minipage}
  \caption{\textit{A posteriori} normalized TKE spectra (left) and PDFs of $\partial u/\partial x$ (right) for Re200 at LES mesh ratio $\mathrm{n}_{\Delta_\mathrm{LR}}=16$. ``DMSRM 200'' and ``SRM 200'' use  the SG-Re200 (multiple) model, while ``DMSRM 200 (NoPUT)'' and ``SRM 200 (NoPUT)'' use the SG-Re200 (multiple) - NoPUT model (see table~\ref{tab:training_configurations}).}
  \label{fig:ModelStability}
\end{figure}

The performance of SRM 200 compared with the reference F-DNS is shown in figure~\ref{fig:ModelStability} using TKE spectra (left) and the PDF of $\partial u/ \partial x$ (right). The results from DMSRM 200 and the DMM are also included. SRM 200 remains numerically stable even without the DSM term and still outperforms the DMM. However, the TKE spectra show a slight accumulation of energy near the cut-off wavenumber, and the PDF of $\partial u/ \partial x$ is narrowed compared to the results obtained using DMSRM 200. Although the DSM contribution in DMSRM 200 is small, these findings suggest that it plays an important role in achieving accurate physical predictions, particularly for wavenumbers near the cut-off filter. 

The stability and accuracy of SRM 200 suggest that the DMSRM's robust performance primarily stems from the ability of the \emph{similarityGAN} framework to reconstruct physically accurate $u^*$ fields, even for unseen $\phi_{\mathrm{LR}}$. To evaluate this aspect, the PUT phase is omitted from the \emph{similarityGAN} training, and the weights and biases obtained are used in both SRM and DMSRM. These models are denoted ``SRM 200 (NoPUT)'' and ``DMSRM 200 (NoPUT).'' As shown in figure~\ref{fig:ModelStability}, both models yield similar predictions to those of implicitly modeled (``no-model'') LES (not shown). Interestingly, the DSM contribution to DMSRM 200 (NoPUT) does not improve the overall performance, indicating that it does not provide enough dissipation under these conditions.  

\begin{figure}[!ht]
  \centering
  \begin{minipage}[b]{0.495\textwidth}
  \includegraphics[width=\textwidth]{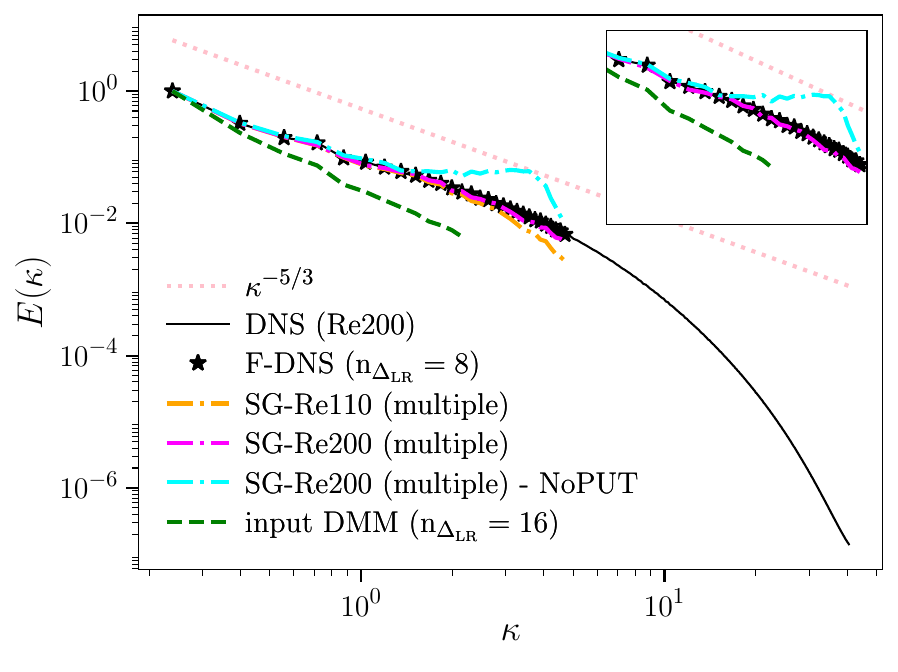}
  \end{minipage}
  \hfill
  \begin{minipage}[b]{0.495\textwidth}
  \centering
  \includegraphics[width=\textwidth]{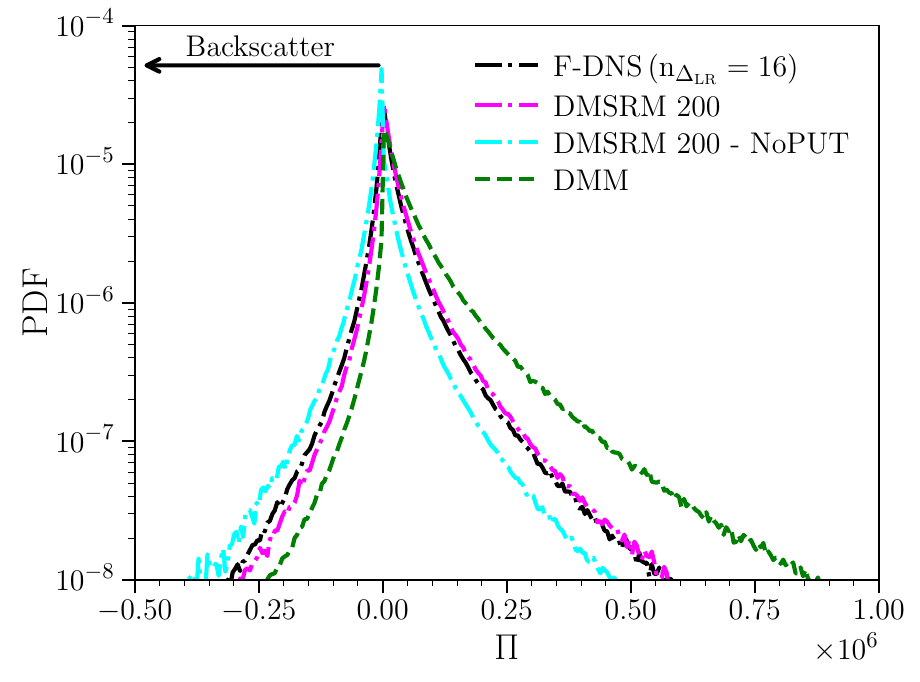}
  \end{minipage}
  \caption{Comparison of the normalized TKE spectra (left) and PDFs of $\Pi$ computed using the super-resolved fields obtained using ``SG-Re110 (multiple)'', ``SG-Re200 (multiple)'', and ``SG-Re200 (multiple) - NoPUT'' models. The same coarse-resolution DMM velocity fields (denoted ``input DMM'') are provided as input velocity fields. The F-DNS and DMM results are reported for reference. ``DMSRM 200 - NoPUT'' refers to the DMSRM using the ``SG-Re200 (multiple) - NoPUT'' model.}
  \label{fig:ModelStability_2}
\end{figure}

To further evaluate the reconstruction capabilities of the \emph{similarityGAN} and the physical fidelity of SR fields generated by the \emph{similarityGAN} from unseen inputs, the same coarse-resolution DMM velocity fields (denoted ``input DMM ($\mathrm{n}_{\Delta_{\mathrm{LR}}}=16$)'') are provided as input to either the SG-Re110, SG-Re200, or SG-Re200 - NoPUT models. As shown in figure~\ref{fig:ModelStability_2} (left), the corresponding $u^*$ fields generated by both PUT-trained models closely match the F-DNS spectra at the corresponding $\mathrm{n}_{\Delta_\mathrm{HR}}$. The small underprediction of TKE near the cut-off of the SG-Re110 model relative to SG-Re200, attributable to the slightly limited generalization capabilities of the SG-Re110 model, explains the small performance degradation of DMSRM 110 observed in the \textit{a posteriori} results. In contrast, omitting the PUT phase entirely leads to a pronounced overprediction of the TKE at the SFS scales. The resulting $u^*$ fields are then unphysical and destabilize the simulations. This is highlighted in figure~\ref{fig:ModelStability_2} (right), where only the DMSRM 200 closely matches the F-DNS across both positive and negative $\Pi$ values. Conversely, in the absence of the PUT phase, the $\Pi$ distribution obtained from DMSRM 200 - NoPUT is unphysical, exhibiting excessive backscatter and insufficient SFS energy dissipation. This imbalance cannot be mitigated by simply adding a dissipative-only DSM term, because the \emph{similarityGAN} model itself produces unphysical $u^*$ fields. In such cases, the on-the-fly stabilization mechanism provided by the DSM term is insufficient to recover a physically meaningful prediction. This suggests that adequate generalization capabilities to unseen input fields are essential for stable and accurate \textit{a posteriori} performance. The PUT phase and training with multiple input resolutions are therefore both essential training steps for approximate filter independence and physically consistent SR reconstruction of the \emph{similarityGAN}, important to achieve stable \textit{a posteriori} LES computations. Nevertheless, even with nearly perfect SR reconstruction, the DSM contribution remains important to obtain correct net energy dissipation.

It must be noted that the aforementioned training steps are not sufficient to ensure generalization to flow configurations fundamentally different from those used for training. Embedding more representative data from diverse flow configurations in the training is expected to be required to extend the practical applicability of the \emph{similarityGAN} framework.


\section{Computational efficiency of the DMSRM on forced HIT}
\label{sec:computationalCost}

In addition to predictive accuracy and stability, computational efficiency is important for the practical deployment of SFS models. Table~\ref{tab:computationa_cost_apriori} summarizes the computational inference costs and memory requirements of the \emph{similarityGAN}'s generator when employed to super-resolve $\phi_{\mathrm{LR}}$ of different resolutions, denoted $N_{\phi_{\mathrm{LR}}}$, corresponding to the LES grid resolutions of Section~\ref{sec:aposteriori_sec}. Two upsampling factors are considered: a factor of two, which is used in the DMSRM, and a larger factor required to reconstruct fields at DNS resolution from the given input resolutions. The larger upsampling factor analysis is included to enable comparison with previous studies, as available works in the literature report analyses upsampled to DNS resolution. For upsampling to DNS resolution, additional upsampling layers are added in sequence in the \emph{similarityGAN} framework (see figure~\ref{fig:similarityGAN}). 

The peak memory requirements and inference times increase significantly with both upsampling factor and $N_{\phi_{\mathrm{LR}}}$. For an upsampling factor of two, memory requirements are modest, and GPU inference provides an order-of-magnitude speed-up compared to the CPUs for both input resolutions, highlighting the suitability of GPUs for accelerating SR-LES workflows with large DL models, such as the \emph{similarityGAN} framework.    

\begin{table}[!ht]
    \centering
    \begin{tabular}{c c c c c c c c}
        \toprule
        \makecell[c]{$N_{\phi_{\mathrm{LR}}}$} & \makecell[c]{Upsampling \\ factor}  & \makecell[c]{Inference \\ memory (peak)} & \multicolumn{2}{c}{\makecell[c]{Inference \\ wall time}} & \multicolumn{3}{c}{\makecell[c]{DMSRM overhead ($\%$ vs DMM)}} \\
        \cmidrule(lr){4-5} \cmidrule(lr){6-8} 
        & & & \makecell[c]{CPUs} &  \makecell[c]{GPU} & Memory (peak) & \makecell[c]{CPU runtime} & \makecell[c]{GPU runtime}  \\
        \midrule
        $32^3$ & 2  & $\approx$0.75 GB & 0.47 s & 0.053 s & 224$\%$ & 102$\%$   & 13$\%$ \\
        $32^3$ & 16 & $\approx$112 GB & 52.27 s & N/A & -- & -- & -- \\
        \midrule
        $64^3$ & 2 & $\approx$3.31 GB &  1.83 s & 0.197 s & 321$\%$ & 123$\%$& 19$\%$\\
        $64^3$ & 8 & $\approx$110 GB & 50.31 s & N/A & -- & -- & --  \\
        \bottomrule
    \end{tabular}
    \caption{Computational inference requirements of \emph{similarityGAN}'s generator for different input field sizes $N_{\phi_{\mathrm{LR}}}$ and upsampling factors. Reported metrics include peak memory usage and wall time on CPU (96-core Intel Sapphire Rapids) and GPU (NVIDIA H100 96 GB). ``N/A'' indicates a memory limitation that prevents execution. All values refer to a single-sample inference. For upsampling factor 2, the \textit{a posteriori} computational overhead of the DMSRM relative to the traditional DMM is also reported, both in terms of memory footprint and runtime cost.}
    \label{tab:computationa_cost_apriori}
\end{table}

The \textit{a posteriori} computational overhead of the DMSRM relative to the DMM is also reported in table~\ref{tab:computationa_cost_apriori} for an upsampling factor of two. Both the runtime and memory requirements are shown. Measurements are averaged over the same 50 timesteps, and the same numerical schemes and pressure solver are used. Initialization routines are excluded from the runtime computations. 

Due to current challenges in parallelizing SR-based LES, all simulations are conducted in serial to avoid potential biases arising from differences in GPU and CPU parallelization strategies \citep{Orland_DL_library}. While an initial analysis of the parallel performance of a similar SR-GAN framework in distributed, inference-coupled LES settings has been reported in Nista \textit{et al.}~\cite{Nista2024_parallel}, efficient parallelization strategies for such LES frameworks remain an open challenge, and no general guidelines are currently available. Furthermore, the relatively small LES domains employed here ($32^3$ and $64^3$) would introduce a disproportionately large halo-to-domain ratio in a parallel setting, making serial execution a better representative and unbiased baseline for overhead assessment. For both the DMM and the DMSRM, one CPU-only execution is performed, while for the DMSRM, an additional setup combining one CPU core and one GPU is also considered. Consequently, none of the LESs employ domain decomposition. 

The DMSRM exhibits considerably larger memory requirements than the DMM (224$\%$ and 321$\%$ overhead for $N_{\phi_{\mathrm{LR}}}=32^3$ and $N_{\phi_{\mathrm{LR}}}=64^3$, respectively), primarily due to the storage of intermediate fields of increased resolution needed by the inference operation. The runtime overhead of the DMSRM is substantial when using CPU-only execution (approximately 100--120$\%$), whereas employing one CPU core and one GPU dramatically reduces the overhead to approximately 13$\%$ and 19$\%$ for the coarse and fine LES grid resolutions, respectively. The differences relative to the DMM arise mainly from CPU–GPU communication, inference, and filtering/downsampling operations at multiple scales (see Eq.~\eqref{eqn:SRSSmodel_toUse}). On GPUs, however, the additional cost is modest, and the overall cost of the DMSRM remains limited. The limited runtime overhead of the DMSRM relative to DMM, particularly when leveraging GPUs, makes it suitable for practical SR-LES applications, especially considering its improved physical accuracy. However, the considerable memory overhead of the DMSRM, arising primarily from the storage of intermediate SR-scale fields, represents the main practical constraint for large-scale deployment.

It must be noted that the memory demand peaks at approximately 110 GB, surpassing the capacity of high-end GPUs such as the NVIDIA H100 (96 GB) for upsampling factors up to DNS resolutions. Consequently, GPU inference is challenging, and CPU inference is slow, exceeding $50~{\rm s}$. This substantial cost escalation is driven by the non-linear scaling of the convolutional operations, activation maps, and intermediate tensors, as well as overhead from the TensorFlow library. These findings emphasize the trade-off between resolution enhancement and computational feasibility of SR-based models and demonstrate that upsampling to DNS resolution is impractical. Decreased upsampling factors are essential to ensure practicality, especially for large-scale SR-LES applications. 

In addition to the \textit{a posteriori} computational overhead discussed above, the one-time training cost of the \emph{similarityGAN} framework should also be considered when evaluating the overall deployment effort. The DMSRM is most beneficial when multiple LES calculations are planned for the same flow configuration, when physical accuracy in terms of backscatter and small-scale intermittency is critical, and when GPU resources are available for inference. The pretrained \emph{similarityGAN} generalizes across Reynolds numbers within the same flow configuration without retraining (see Sections~\ref{sec:apriori_sec} and~\ref{sec:aposteriori_sec}). However, the model has been assessed only on forced HIT, and its transfer to fundamentally different flow configurations without retraining is not demonstrated in this work and warrants additional investigations. These trade-offs should be carefully considered by potential adopters of the DMSRM formulation.

\section{Conclusion}

A novel hybrid dynamic super-resolution-based mixed model is presented, which combines the robustness and computational efficiency of dynamic mixed models with the physically consistent velocity reconstructions of SR-GAN frameworks, here obtained using \emph{similarityGAN}. In the DMSRM, the velocity fields are super-resolved to finer-than-LES resolutions, and both the scale-similarity term and the dynamic Smagorinsky term are evaluated using the resulting inter-scale range. The physical fidelity of the resulting SFS stress closure is improved compared to conventional DMMs without the prohibitive need to upsample to DNS-scale resolutions as in previous SR-based closure approaches.

\textit{A priori} analyses on forced homogeneous isotropic turbulence demonstrate that the \emph{similarityGAN} framework accurately reconstructs super-resolved velocity fields, even for ``out-of-sample'' filter conditions. Training with multiple input resolutions is essential for maintaining accuracy across varying input filter sizes, and the framework is also shown to generalize well to higher Reynolds number flows within the same flow configuration. The scale-similarity term evaluated at the SR-grid level accurately reproduces pointwise and statistical features of the DNS SFS stress tensor, whereas traditional scale-similarity models tend to overpredict the magnitude of the stresses and backscatter. The DSM contribution is minor in \textit{a priori} tests, suggesting that the super-resolution component alone could serve as an accurate closure model.

\textit{A posteriori} tests confirm that the DMSRM improves the predictive capability and physical fidelity relative to the DMM for different LES resolutions in forced HIT. For coarse LES, the DMSRM accurately captures the TKE spectra, velocity gradient PDFs, and small-scale intermittency. Analyses of SFS energy dissipation show that the DMSRM predicts correct levels of backscatter, unlike the DMM, which tends to underestimate backscatter and overpredict SFS dissipation. No strong sensitivity to the Reynolds number of the training dataset is observed for the DMSRM, suggesting adequate generalization capabilities within the same flow configurations. For finer LES, differences between models are less pronounced, yet the DMSRM consistently improves upon the DMM by preserving small-scale structures and avoiding excessive dissipation. Under the configurations tested, the computational costs of the DMSRM are also manageable, as the moderate upsampling factors employed by the DMSRM keep inference efficient.

These investigations further demonstrate that an accurate scale-similarity contribution is essential for achieving close agreement with F-DNS, while the DSM term plays a secondary role by ensuring energy balance at the cut-off scale and exerts a stabilizing effect. The superior performance of the DMSRM primarily stems from the physically consistent super-resolved velocity fields generated by \emph{similarityGAN} and its generalization to unseen inputs of forced HIT. This capability is enabled by the partially unsupervised training phase, in which the discriminator functions as a physics-informed component and the generator is guided through adversarial feedback to produce physically plausible SR fields, even for unseen input fields. The SFS stress tensor estimation at the grid-filter scale is then obtained directly from such SR fields using an explicit filter kernel and size consistent with training, yielding stresses of the correct magnitude and an accurate reproduction of the dissipation.

Considering the forced HIT datasets employed, the DMSRM achieves improved physical fidelity at a moderate runtime overhead when GPUs are available, making it a robust and practical alternative to traditional DMMs for LES. Although the present study focuses on homogeneous isotropic turbulence, future work will be devoted to extending the DMSRM to inhomogeneous flows. Moreover, extending the DMSRM formulation to variable-density flows and applying it to wall-bounded turbulent flows might be of interest, where additional physical constraints may be required to ensure physically consistent SR velocity and scalar fields. Furthermore, assessing and improving the generalization capabilities of SR-GAN frameworks across different flow configurations will be essential for practical deployment.

\section*{Acknowledgments}
The research leading to these results has received funding from the German Federal Ministry of Education and Research (BMBF) and the state of North Rhine-Westphalia for supporting this work as part of the NHR funding and the European Union under the European Research Council Advanced Grant HYDROGENATE, Grant Agreement No.~\textit{101054894}. The authors thank S.~Sakhare for his exceptional support and contribution to this research project. Computations were performed with computing resources granted by RWTH Aachen University under project \textit{rwth1387}.

\section*{Declaration of interests}
The authors report no conflict of interest.


\appendix

\section{Definition of filter kernel and downsampling operators}
\label{sec:appendinx_filters}

The definitions of the spatial box, Gaussian, spectral, and modified-Gaussian (MG) filter kernels are
\begin{align}
    \mathcal{G}_\mathrm{box}({\boldsymbol{r}})&={\begin{cases}{\frac{1}{\Delta}} \, {\qquad \text{if}}\left|{\boldsymbol{r}}\right|\leq{\frac{\Delta }{2}} \\0 \, {\qquad \text{otherwise}}\end{cases}} 
    \\
    \mathcal{G}_\mathrm{Gaussian}({\boldsymbol{r}})&=\left({\frac{6}{\pi \Delta ^{{2}}}}\right)^{{{\frac  {1}{2}}}}\exp {\left(-{\frac  {6{\boldsymbol  {r}}^{2}}{\Delta ^{2}}}\right)},
    \\
    \mathcal{G}_\mathrm{spectral}({\boldsymbol  {r}})&={\frac  {\sin {(\pi {\boldsymbol  {r}}/\Delta )}}{\pi {\boldsymbol  {r}} }}
    \\
    \mathcal{G}_\mathrm{MG}(\boldsymbol{r}) &=
\left( \frac{\beta}{\pi \Delta^2} \right)^{1/2} \exp\left( -\frac{\beta \boldsymbol{r}^2}{\Delta^2} \right) \, .
\label{eqn:filter_kernels}
\end{align}
The kernels are given in one dimension and applied successively in each spatial direction. Here, the filter width $\Delta$ is set to either $\Delta_{\mathrm{LR}}$ or $\Delta_{\mathrm{HR}}$, depending on whether the $\phi_{\mathrm{LR}}$ or $\phi_{\mathrm{HR}}$ fields are considered, as detailed in Section~\ref{sec:dataset}. The filter operator in Eq.~\eqref{eqn:filter_kernels} is applied to generate the ``out-of-sample'' $\phi_{\mathrm{LR}}$ fields used for the \textit{a priori} tests in Section~\ref{sec:apriori_sec}. In particular, for the MG kernel, the parameter $\beta=3$ is chosen to introduce TKE distortion near the cut-off filter scale. This ensures that the test fields differ systematically from those generated with the box, Gaussian, and spectral filter kernels used to generate the training $\phi_{\mathrm{LR}}$ fields (see Section~\ref{sec:dataset}). The downsampling operator, applied independently of the filter kernel, maps the velocity field from $\Omega^*$ to $\overline{\Omega}$ by simply subsampling. For a decimation factor of two, it reads
\begin{equation}
\Check{\mathbf{u}}(I,J,K) = \mathbf{u}(2I,\, 2J,\, 2K), \quad
I = 0, \dots, \overline{N}_x-1,\;\;
J = 0, \dots, \overline{N}_y-1,\;\;
K = 0, \dots, \overline{N}_z-1,
\end{equation}
where $\Check{\mathbf{u}}$ is the downsampled velocity field and $\overline{N}_x, \overline{N}_y, \overline{N}_z$ are the number of coarse-grid points in each direction (see table~\ref{tab:filtering_description}).

\section{Calculation of the dynamic model coefficient for the DMM}
\label{sec:DMM_Cd_definition}

The DMM used for comparison is defined as
 \begin{equation}
    \tau^{\mathrm{DMM}}_{ij} = -2 C_{d}^2 \overline{\Delta}^2 |\overline{S}| \overline{S}_{ij} + \tau^{\mathrm{SSM}}_{ij} \, ,
     \label{eqn:DMM_2} 
 \end{equation}
where $\overline{S}_{ij}$ and $|\overline{S}|$ are defined in Eq.~\eqref{eqn:strain_rate}, and $\tau^{\mathrm{SSM}}_{ij}$ in Eq.~\eqref{eqn:SSM}. The dynamic model coefficient $C_d^2$ is obtained by substituting Eq.~\eqref{eqn:DMM_2} into the Germano identity, analogous to Eq.~\eqref{eqn:Germano_identity}, giving
\begin{align}
    H_{ij} + C^2_d M_{ij} = L_{ij} \, \mathrm{with} \, & \nonumber \, L_{ij} = \widehat{\overline{u}_{i} \overline{u}_{j}} - \widehat{\overline{u}}_{i} \widehat{\overline{u}}_{j} \\
    & M_{ij} = -2\overline{\Delta}^2 (4 |\widehat{\overline{S}}| \widehat{\overline{{S}}}_{ij} - \widehat{|\overline{S}| \, \overline{S}}_{ij}) \, .
\end{align}
The tensor $H_{ij}$ is defined as
\begin{equation}
    H_{ij} = \widetilde{\widehat{\overline{u}_{i}} \widehat{\overline{u}}_{j}} - \widetilde{\widehat{\overline{u}}}_{i} \widetilde{\widehat{\overline{u}}}_{j} - (\widehat{\widehat{\overline{u}_{i} \overline{u}_{j}}} - \widehat{\widehat{\overline{u}_{i}}\,\widehat{\overline{u}_{j}}}) \, ,
    \label{eqn:H_ij_DMM}
\end{equation}
with notations and operators defined in table~\ref{tab:filtering_description}.  

The dynamic model coefficient $C^2_d$ is then computed using a least-squares error approach, such that
\begin{equation}
    C^2_d = \frac{\langle M_{ij}(L_{ij} - H_{ij}) \rangle}{\langle M_{ij}M_{ij} \rangle} \, ,
    \label{eqn:C_d_DMM}
\end{equation}
where $\langle \cdot \rangle$ denotes averaging over homogeneous directions to ensure numerical stability. Definitions in Eqs.~\eqref{eqn:H_ij_DMM} and~\eqref{eqn:C_d_DMM} might differ from prior works, e.g., in \cite{zang_mixed, salvetti1995priori, vreman1994formulation, anderson1999effects}, due to differences in the SSM definition, the number of dynamic parameters involved, and operators employed among the different filter scales.

\section{On the \textit{a priori} correlation of the SSM}
\label{sec:correlation_SS}

Previous studies have reported Pearson correlation coefficients between the modeled SFS stress tensor, $\tau^{\mathrm{SSM}}$, and the true stress tensor obtained from DNS data, $\tau^{\mathrm{F\mbox{-}DNS}}$, often exceeding 0.7~\citep{liu1994properties, liu1995experimental, sarghini1999scale}. However, the Pearson correlation coefficients obtained in this work and reported in figure~\ref{fig:jPDF_apriori_Delta16} are significantly lower, ranging from approximately 0.6 to 0.5, depending on $\mathrm{n_{\Delta_{LR}}}$ used.

\begin{figure}[!ht]
     \centering
     \includegraphics[width=0.5\linewidth]{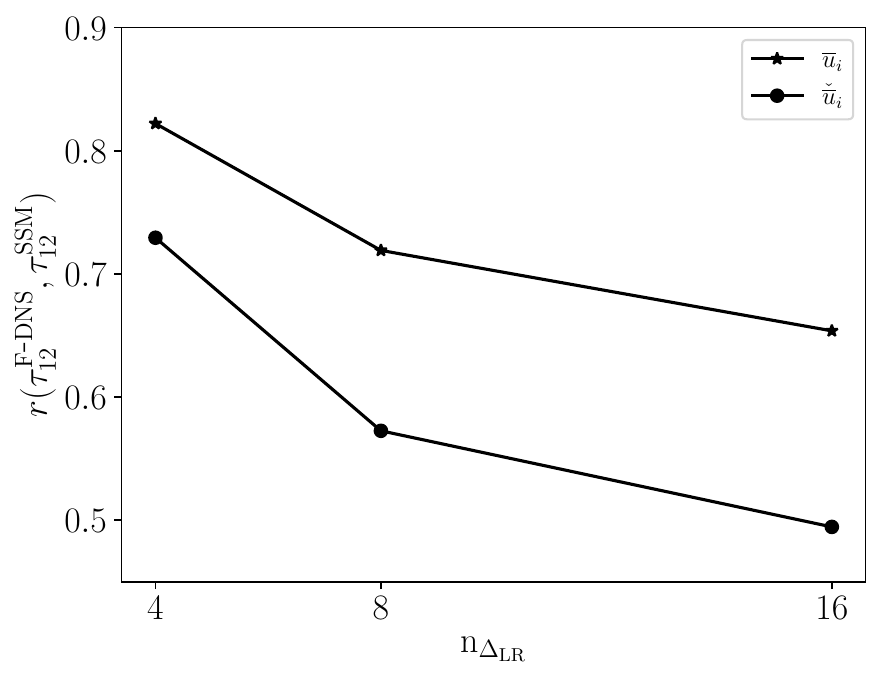}
     \caption{Pearson correlation coefficient $r$ between the true stress tensor component, $\tau^{\mathrm{F\mbox{-}DNS}}_{12}$, and the corresponding SSM component, $\tau^{\mathrm{SSM}}_{12}$, evaluated across different input $\phi_{\mathrm{LR}}$ resolutions for two preprocessing approaches: explicit spatial filtering only, $\overline{u}_{i}$, and explicit filtering with subsequent downsampling consistent with the filter width, $\check{\overline{u}}_{i}$.}
     \label{fig:correlation_SS}
\end{figure}

Figure~\ref{fig:correlation_SS} shows the Pearson correlation coefficient between $\tau_{12}^{\mathrm{SSM}}$ and $\tau_{12}^{\mathrm{F\mbox{-}DNS}}$ for two preprocessing approaches used to obtain $\overline{u}_i$, evaluated across different input $\phi_{\mathrm{LR}}$ resolutions. In most \textit{a priori} evaluations, $\overline{u}_i$ is obtained by applying only an explicit spatial filter to the DNS velocity, without subsequent downsampling. This approach retains spectral content that would not be present in an actual LES and leads to high correlation values, as shown in figure~\ref{fig:correlation_SS}. Such an approach is typically adopted when the objective is to verify physical hypotheses or mathematical properties of a model, rather than assess its practical LES performance. In contrast, as employed in this work, when both explicit filtering and corresponding downsampling are applied, $\check{\overline{u}}_i$, a notable reduction of approximately 20 to 25$\%$ in Pearson correlation coefficients is observed, as shown in figure~\ref{fig:correlation_SS}. Therefore, this reduction in Pearson correlation coefficient arises mainly from differences in how the input grid (LES) velocity field is obtained in \textit{a priori} settings.


\section{Influence of numerical discretization on the DMSRM formulation}
\label{sec:appendix_discretization}

\textit{A posteriori} LES simulations are performed using a second-order finite-difference scheme for spatial discretization, compared to the fourth-order finite-difference scheme employed throughout the main body of this work, to assess the influence of the numerical discretization method on the DMSRM formulation. For this analysis, only the DMSRM 200 is considered, and the DMM with the fourth-order discretization scheme is reported as a baseline.

\begin{figure}[!ht]
  \centering
  \begin{minipage}[b]{0.495\textwidth}
  \includegraphics[width=\textwidth]{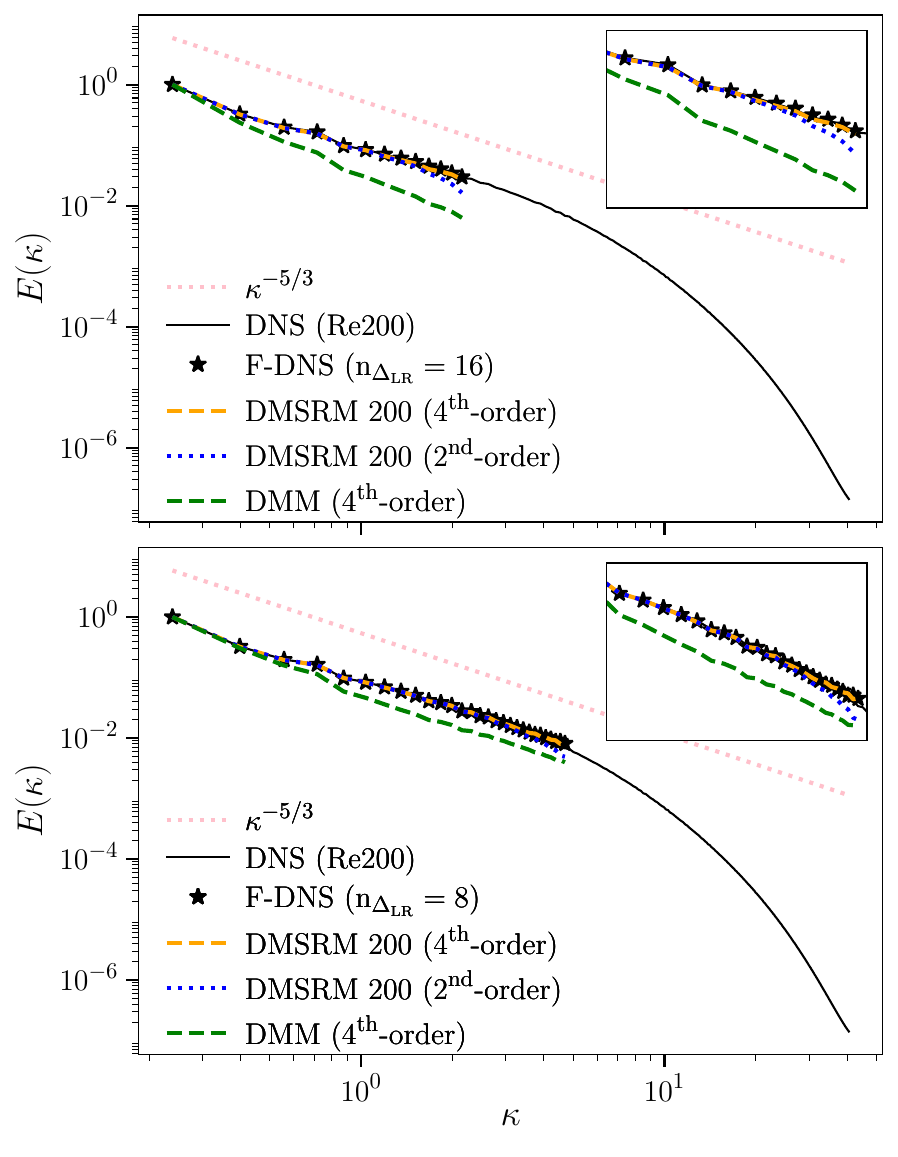}
  \end{minipage}
  \hfill
  \begin{minipage}[b]{0.495\textwidth}
  \centering
  \includegraphics[width=\textwidth]{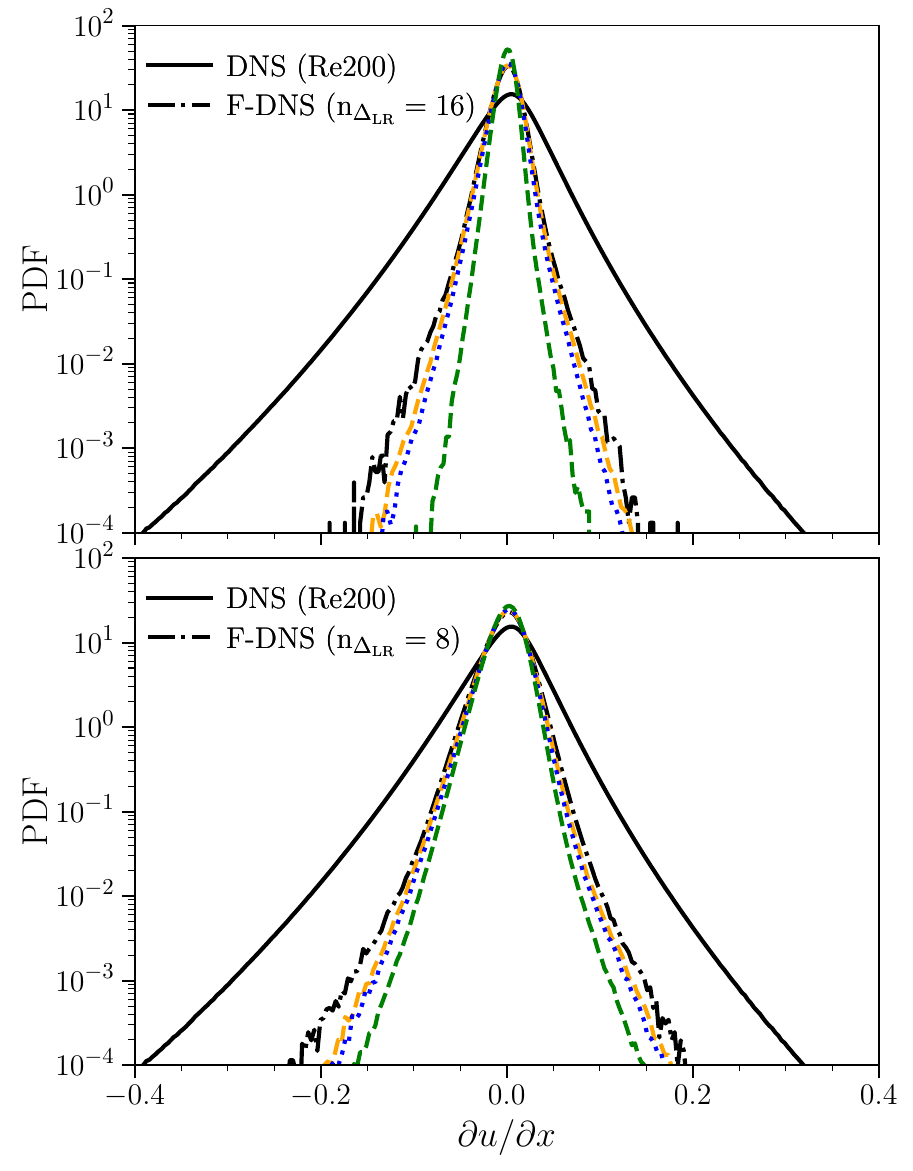}
  \end{minipage}
  \caption{\textit{A posteriori} normalized TKE spectra (left) and PDFs of $\partial u/\partial x$ (right) of LES coarsened by $\mathrm{n}_{\Delta_\mathrm{LR}}=16$ (top row) and $\mathrm{n}_{\Delta_\mathrm{LR}}=8$ (bottom row) from the Re200 DNS, comparing DMSRM predictions using fourth-order and second-order finite-difference discretizations with the DMM using fourth-order discretization and the F-DNS reference. ``DMSRM 200'' employs the SG-Re200 (multiple) model in both cases.}
  \label{fig:TKE_PDF_aposteriori_differentDiscretization}
\end{figure}

Figure~\ref{fig:TKE_PDF_aposteriori_differentDiscretization} shows the TKE spectra (left) and PDFs of $\partial u/\partial x$ (right) for the DMSRM 200 with both fourth- and second-order discretizations, compared with the DMM with the fourth-order discretization only, at two LES resolutions ($\mathrm{n}_{\Delta_\mathrm{LR}} = 16$ and $\mathrm{n}_{\Delta_\mathrm{LR}} = 8$). The DMSRM 200 outperforms the DMM regardless of the discretization order employed for the DMSRM. However, a small TKE underprediction near the grid cut-off is observed when switching from fourth- to second-order discretization, which is slightly more pronounced at the coarser resolution. This is consistent with the additional numerical dissipation introduced by the second-order scheme near the cut-off wavenumber. Including LES fields generated with second-order discretization schemes during the PUT phase might further improve the robustness of the framework to different numerical schemes, and is left for future work.

\vspace{1em}

\bibliographystyle{IEEEtran}  
\bibliography{sample}

@MISC{Nista_datapublication,
      author       = {Nista, Ludovico and Schumann, Christoph David Karl and
                      Vivenzo, Marco and Fröde, Fabian and Grenga, Temistocle and
                      MacArt, Jonathan F. and Attili, Antonio and Pitsch, Heinz},
      title        = {{H}omogeneous isotropic turbulence database for training
                      super-resolution data-driven turbulence closure models},
      reportid     = {RWTH-2024-03259},
      year         = {2024},
      doi          = {10.18154/RWTH-2024-03259},
    howpublished = {https://doi.org/10.18154/RWTH-2024-03259}
}

@misc{kingma2017adammethodstochasticoptimization,
      title={Adam: A Method for Stochastic Optimization}, 
      author={Diederik P. Kingma and Jimmy Ba},
      year={2017},
      eprint={1412.6980},
      archivePrefix={arXiv},
      primaryClass={cs.LG},
      url={https://arxiv.org/abs/1412.6980}, 
}

@article{pitschReviewLES,
   author = {Pitsch, Heinz},
   title = {Large-eddy simulation of turbulent combustion}, 
   journal= {Annual Review of Fluid Mechanics},
   year = {2006},
   volume = {38},
   number = {1},
   pages = "453--482",
   doi = "https://doi.org/10.1146/annurev.fluid.38.050304.092133",
   url = "https://www.annualreviews.org/content/journals/10.1146/annurev.fluid.38.050304.092133",
   publisher = {Annual Reviews},
   issn = "1545-4479"
  }

@article{goodfellow2020generative,
  title={Generative adversarial networks},
  author={Goodfellow, Ian and Pouget-Abadie, Jean and Mirza, Mehdi and Xu, Bing and Warde-Farley, David and Ozair, Sherjil and Courville, Aaron and Bengio, Yoshua},
  journal={Communications of the ACM},
  volume={63},
  number={11},
  pages={139--144},
  year={2020},
  publisher={ACM New York, NY, USA}
}

@inproceedings{wang2018esrgan,
  author    = {Wang, Xintao and Yu, Ke and Wu, Shixiang and Gu, Jinjin and Liu, Yihao and Dong, Chao and Qiao, Yu and Loy, Chen Change},
  title     = {{ESRGAN}: Enhanced super-resolution generative adversarial networks},
  booktitle = {Proceedings of the European Conference on Computer Vision Workshops},
  publisher = {Springer},
  pages     = {63--79},
  year      = {2018},
  doi       = {10.1007/978-3-030-11021-5_5}
}

@article{arumapperuma2025extrapolation,
  title   = {Extrapolation performance of convolutional neural network-based combustion models for large-eddy simulation: influence of {R}eynolds number, filter kernel and filter size},
  author  = {Arumapperuma, Geveen and Sorace, Nicola and Jansen, Matthew and Bladek, Oliver and Nista, Ludovico and Sakhare, Shreyans and Berger, Lukas and Pitsch, Heinz and Grenga, Temistocle and Attili, Antonio},
  journal = {Flow, Turbulence and Combustion},
  volume  = {115},
  number  = {3},
  pages   = {1261--1290},
  year    = {2025},
  doi     = {10.1007/s10494-025-00643-w}
}

@misc{jolicoeur2018relativistic,
  title        = {The relativistic discriminator: a key element missing from standard {GAN}},
  author       = {Jolicoeur-Martineau, Alexia},
  year         = {2018},
  howpublished = {arXiv:1807.00734}
}

@INPROCEEDINGS{ledig2017photo,
  author={Ledig, Christian and Theis, Lucas and Huszár, Ferenc and Caballero, Jose and Cunningham, Andrew and Acosta, Alejandro and Aitken, Andrew and Tejani, Alykhan and Totz, Johannes and Wang, Zehan and Shi, Wenzhe},
  booktitle={2017 IEEE Conference on Computer Vision and Pattern Recognition}, 
  title={Photo-Realistic Single Image Super-Resolution Using a Generative Adversarial Network}, 
  year={2017},
  volume={},
  number={},
  pages={105-114},
  doi={10.1109/CVPR.2017.19}
}

@phdthesis{bardina1983improved,
  title  = {Improved turbulence models based on large eddy simulation of homogeneous, incompressible, turbulent flows},
  author = {Bardina, Jorge},
  school = {Stanford University},
  year   = {1983}
}

@article{kim2021unsupervised,
  title   = {Unsupervised deep learning for super-resolution reconstruction of turbulence},
  author  = {Kim, Hyojin and Kim, Junhyuk and Won, Sungjin and Lee, Changhoon},
  journal = {Journal of Fluid Mechanics},
  volume  = {910},
  pages   = {A29},
  year    = {2021},
  doi     = {10.1017/jfm.2020.1028}
}

@article{gamahara2017searching,
  title={Searching for turbulence models by artificial neural network},
  author={Gamahara, Masataka and Hattori, Yuji},
  journal={Physical Review Fluids},
  volume={2},
  number={5},
  pages={054604},
  year={2017},
  publisher={APS}
}

@article{lilly1992proposed,
  title={A proposed modification of the Germano subgrid-scale closure method},
  author={Lilly, Douglas K},
  journal={Physics of Fluids A: Fluid Dynamics},
  volume={4},
  number={3},
  pages={633--635},
  year={1992},
  publisher={American Institute of Physics}
}

@inproceedings{nista2021turbulent,
  title={Turbulent mixing predictive model with physics-based Generative Adversarial Network},
  author={Nista, Ludovico and Schumann, Christoph David Karl and Grenga, Temistocle and Karimi, Amir N. and Scialabba, Gandolfo and Bode, Mathis and Attili, Antonio and Pitsch, Heinz},
  booktitle={10th European Combustion Meeting},
  pages={460--465},
  year={2021}
}

@article{desjardins2008high,
  title={High order conservative finite difference scheme for variable density low {M}ach number turbulent flows},
  author={Desjardins, Olivier and Blanquart, Guillaume and Balarac, Guillaume and Pitsch, Heinz},
  journal={Journal of Computational Physics},
  volume={227},
  year={2008},
  publisher={Elsevier}
}

@article{vinuesa2021potential,
  title   = {Enhancing computational fluid dynamics with machine learning},
  author  = {Vinuesa, Ricardo and Brunton, Steven L.},
  journal = {Nature Computational Science},
  volume  = {2},
  number  = {6},
  pages   = {358--366},
  year    = {2022},
  doi     = {10.1038/s43588-022-00264-7}
}

@article{IHME2022101010,
title = {Combustion machine learning: Principles, progress and prospects},
journal = {Progress in Energy and Combustion Science},
volume = {91},
pages = {101010},
year = {2022},
author = {Matthias Ihme and Wai Tong Chung and Aashwin Ananda Mishra}
}

@InProceedings{10.1007/978-3-030-44584-3_43,
author="von Rueden, Laura
and Mayer, Sebastian
and Sifa, Rafet
and Bauckhage, Christian
and Garcke, Jochen",
editor="Berthold, Michael R.
and Feelders, Ad
and Krempl, Georg",
title="Combining Machine Learning and Simulation to a Hybrid Modelling Approach: Current and Future Directions",
booktitle="Advances in Intelligent Data Analysis XVIII",
year="2020",
publisher="Springer International Publishing",
address="Cham",
pages="548--560",
isbn="978-3-030-44584-3"
}

@InProceedings{GAN_lr,
  title = 	 {Which Training Methods for {GAN}s do actually Converge?},
  author =       {Mescheder, Lars and Geiger, Andreas and Nowozin, Sebastian},
  booktitle = 	 {Proceedings of the 35th International Conference on Machine Learning},
  pages = 	 {3481--3490},
  year = 	 {2018},
  editor = 	 {Dy, Jennifer and Krause, Andreas},
  volume = 	 {80},
  series = 	 {Proceedings of Machine Learning Research},
  month = 	 {10--15 Jul},
  publisher =    {Proceedings of Machine Learning Research},

}

@book{sagaut2006large,
  title={Large Eddy Simulation for Incompressible Flows: An Introduction},
  author={Sagaut, Pierre and Meneveau, Charles},
  series={Scientific Computation},
  year={2006},
  number={7},
  publisher={Springer}
}

@article{fractalInterpolation,
  title = {Fractal Model for Coarse-Grained Nonlinear Partial Differential Equations},
  author = {Scotti, Alberto and Meneveau, Charles},
  journal = {Physical Review Letters},
  volume = {78},
  issue = {5},
  pages = {867--870},
  numpages = {0},
  year = {1997},
  month = {Feb},
  publisher = {American Physical Society},
  doi = {10.1103/PhysRevLett.78.867},
  url = {https://link.aps.org/doi/10.1103/PhysRevLett.78.867}
}

@article{zang_mixed,
author = {Zang,Yan  and Street,Robert L.  and Koseff,Jeffrey R. },
title = {A dynamic mixed subgrid‐scale model and its application to turbulent recirculating flows},
journal = {Physics of Fluids A: Fluid Dynamics},
volume = {5},
number = {12},
pages = {3186-3196},
year = {1993}   
}

@article{berger2025combustion,
  title={{LES} combustion model for premixed turbulent hydrogen flames with thermodiffusive instabilities: a priori and a posteriori analysis},
  author={Berger, Lukas and Attili, Antonio and Gauding, Michael and Pitsch, Heinz},
  journal={Journal of Fluid Mechanics},
  volume={1003},
  pages={A33},
  year={2025},
  publisher={Cambridge University Press}
}

@article{sarghini1999scale,
  title={Scale-similar models for large-eddy simulations},
  author={Sarghini, Fabrizio and Piomelli, Ugo and Balaras, Elias},
  journal={Physics of Fluids},
  volume={11},
  number={6},
  pages={1596--1607},
  year={1999},
  publisher={American Institute of Physics}
}

@incollection{piomelli1996large,
  title={Large-eddy simulations: theory and applications},
  author={Piomelli, Ugo and Chasnov, Jeffrey Robert},
  booktitle={Turbulence and Transition Modelling: Lecture Notes from the ERCOFTAC/IUTAM Summerschool held in Stockholm, 12--20 June, 1995},
  pages={269--336},
  year={1996},
  publisher={Springer}
}

@article{meneveau2000scale,
  title={Scale-invariance and turbulence models for large-eddy simulation},
  author={Meneveau, Charles and Katz, Joseph},
  journal={Annual Review of Fluid Mechanics},
  volume={32},
  number={1},
  pages={1--32},
  year={2000},
  publisher={Annual Reviews 4139 El Camino Way, PO Box 10139, Palo Alto, CA 94303-0139, USA}
}

@inproceedings{Orland_DL_library,
  author    = {Orland, Fabian and Nista, Ludovico and Kocher, Nick and Vanvinckenroye, Joris and Pitsch, Heinz and Terboven, Christian},
  title     = {Efficient and scalable {AIxeleration} of reactive {CFD} solvers coupled with deep learning inference on heterogeneous architectures},
  booktitle = {Proceedings of the 2025 International Conference on High Performance Computing in Asia-Pacific Region Workshops (HPC Asia '25 Workshops)},
  publisher = {Association for Computing Machinery},
  address   = {New York, NY, USA},
  pages     = {45--57},
  year      = {2025},
  doi       = {10.1145/3703001.3724386}
}

@article{fukami2023super,
  title   = {Super-resolution analysis via machine learning: a survey for fluid flows},
  author  = {Fukami, Kai and Fukagata, Koji and Taira, Kunihiko},
  journal = {Theoretical and Computational Fluid Dynamics},
  volume  = {37},
  number  = {4},
  pages   = {421--444},
  year    = {2023},
  doi     = {10.1007/s00162-023-00663-0}
}

@article{germano1991dynamic,
  title = {A dynamic subgrid‐scale eddy viscosity model},
  author = {Germano, Massimo and Piomelli, Ugo and Moin, Parviz and Cabot, William H.},
  journal={Physics of Fluids A: Fluid Dynamics},
  volume={3},
  number={7},
  pages={1760--1765},
  year={1991},
  month = {07},
  publisher={American Institute of Physics},
  doi = {10.1063/1.857955},
}

@article{durbin2018some,
  title={Some recent developments in turbulence closure modeling},
  author={Durbin, Paul A},
  journal={Annual Review of Fluid Mechanics},
  volume={50},
  pages={77--103},
  year={2018},
  publisher={Annual Reviews}
}

@article{nista2024PRF,
  title = {Influence of adversarial training on super-resolution turbulence reconstruction},
  author = {Nista, Ludovico and Pitsch, Heinz and Schumann, Christoph David Karl and Bode, Mathis and Grenga, Temistocle and MacArt, Jonathan F. and Attili, Antonio},
  journal = {Physical Review Fluids},
  volume = {9},
  issue = {6},
  pages = {064601},
  numpages = {24},
  year = {2024},
  month = {Jun},
  publisher = {American Physical Society},
  doi = {10.1103/PhysRevFluids.9.064601}
}

@article{sirignano2020,
  title={{DPM}: A deep learning {PDE} augmentation method 
         with application to large-eddy simulation},
  author={Sirignano, Justin and MacArt, Jonathan F and Freund, 
          Jonathan B},
  journal={Journal of Computational Physics},
  volume={423},
  pages={109811},
  year={2020}
}

@article{wu2025generalizable,
  title={Generalizable super-resolution turbulence reconstruction from minimal training data},
  author={Wu, Haokai and Cao, Yong and Chen, Yaoran and Laima, Shujin and Chen, Wen-Li and Zhou, Dai and Li, Hui},
  journal={Journal of Fluid Mechanics},
  volume={1024},
  pages={A27},
  year={2025},
  publisher={Cambridge University Press}
}

@article{attili2012statistics,
  title={Statistics and scaling of turbulence in a spatially developing mixing layer at {$Re_{\lambda}$}= 250},
  author={Attili, Antonio and Bisetti, Fabrizio},
  journal={Physics of Fluids},
  volume={24},
  number={3},
  year={2012},
  pages = {035109},
  publisher={AIP Publishing}
}

@article{beck2019deep,
  title={Deep neural networks for data-driven {LES} closure models},
  author={Beck, Andrea and Flad, David and Munz, Claus-Dieter},
  journal={Journal of Computational Physics},
  volume={398},
  pages={108910},
  year={2019},
  publisher={Elsevier}
}

@article{kurz2023deep,
  title={Deep reinforcement learning for turbulence modeling 
         in large eddy simulations},
  author={Kurz, Marius and Offenh{\"a}user, Philipp and Beck, Andrea},
  journal={International Journal of Heat and Fluid Flow},
  volume={99},
  pages={109094},
  year={2023}
}

@article{novati2021automating,
  title={Automating turbulence modelling by multi-agent reinforcement learning},
  author={Novati, Guido and De Laroussilhe, Hugues Lascombes and Koumoutsakos, Petros},
  journal={Nature Machine Intelligence},
  volume={3},
  number={1},
  pages={87--96},
  year={2021},
  publisher={Nature Publishing Group UK London}
}

@article{macart_adjoint,
  title = {Embedded training of neural-network subgrid-scale turbulence models},
  author = {MacArt, Jonathan F. and Sirignano, Justin and Freund, Jonathan B.},
  journal = {Physical Review Fluids},
  volume = {6},
  issue = {5},
  pages = {050502},
  numpages = {28},
  year = {2021},
  month = {May},
  publisher = {American Physical Society},
  doi = {10.1103/PhysRevFluids.6.050502},
  url = {https://link.aps.org/doi/10.1103/PhysRevFluids.6.050502}
}

@article{Morinishi_coupledDMM,
    author = {Morinishi, Youhei and Vasilyev, Oleg V.},
    title = {A recommended modification to the dynamic two-parameter mixed subgrid scale model for large eddy simulation of wall bounded turbulent flow},
    journal = {Physics of Fluids},
    volume = {13},
    number = {11},
    pages = {3400-3410},
    year = {2001},
    month = {11},
    doi = {10.1063/1.1404396}
}

@article{Nista2024_parallel,
title = {Parallel implementation and performance of super-resolution generative adversarial network turbulence models for large-eddy simulation},
journal = {Computers \& Fluids},
volume = {288},
pages = {106498},
year = {2025},
issn = {0045-7930},
doi = {https://doi.org/10.1016/j.compfluid.2024.106498},
author = {Ludovico Nista and Christoph David Karl Schumann and Peicho Petkov and Valentin Pavlov and Temistocle Grenga and Jonathan F. MacArt and Antonio Attili and Stoyan Markov and Heinz Pitsch}
}

@article{eswaran1988examination,
  title={An examination of forcing in direct numerical simulations of turbulence},
  author={Eswaran, Vinayak and Pope, Stephen B},
  journal={Computers \& Fluids},
  volume={16},
  number={3},
  pages={257--278},
  year={1988},
  publisher={Elsevier}
}

@article{liu1994properties,
  title={On the properties of similarity subgrid-scale models as deduced from measurements in a turbulent jet},
  author={Liu, Shewen and Meneveau, Charles and Katz, Joseph},
  journal={Journal of Fluid Mechanics},
  volume={275},
  pages={83--119},
  year={1994},
  publisher={Cambridge University Press}
}

@article{anderson1999effects,
  title={Effects of the similarity model in finite-difference {LES} of isotropic turbulence using a {L}agrangian dynamic mixed model},
  author={Anderson, Richard and Meneveau, Charles},
  journal={Flow, Turbulence and Combustion},
  volume={62},
  pages={201--225},
  year={1999},
  publisher={Springer}
}

@article{vreman1994formulation,
  title={On the formulation of the dynamic mixed subgrid-scale model},
  author={Vreman, Bert and Geurts, Bernard and Kuerten, Hans},
  journal={Physics of Fluids},
  volume={6},
  number={12},
  pages={4057--4059},
  year={1994},
  publisher={American Institute of Physics}
}

@article{salvetti1995priori,
  title={A priori tests of a new dynamic subgrid-scale model for finite-difference large-eddy simulations},
  author={Salvetti, Maria Vittoria and Banerjee, Sanjoy},
  journal={Physics of Fluids},
  volume={7},
  number={11},
  pages={2831--2847},
  year={1995},
  publisher={American Institute of Physics}
}

@article{liu1995experimental,
  title={Experimental study of similarity subgrid-scale models of turbulence in the far-field of a jet},
  author={Liu, Shewen and Meneveau, Charles and Katz, Joseph},
  journal={Applied Scientific Research},
  volume={54},
  pages={177--190},
  year={1995},
  doi={10.1007/BF00849115},
  publisher={Springer}
}

@article{mcdonough1998data,
  title = {A data-fitting procedure for chaotic time series},
  author={McDonough, James M. and Mukerji, Sudip and Chung, Serena H.},
  journal = {Applied Mathematics and Computation},
  volume={95},
  number={2-3},
  pages={219--243},
  year={1998},
  doi = {https://doi.org/10.1016/S0096-3003(97)10062-5},
  publisher={Elsevier}
}

@article{domaradzki2002large,
  title={Large eddy simulations using the subgrid-scale estimation model and truncated {N}avier--{S}tokes dynamics},
  author={Domaradzki, J Andrzej and Loh, Kuo Chieh and Yee, Patrick P},
  journal={Theoretical and Computational Fluid Dynamics},
  volume={15},
  pages={421--450},
  year={2002},
  publisher={Springer}
}

\end{document}